%% file: main-tosem.tex
\documentclass[acmtosem,screen]{acmart}

\input{common/preamble}
\input{common/commands}

\setcopyright{acmcopyright}

\begin{document}

\title[\papertitleshort]{\papertitle}

\author{\authorBname}
\affiliation{%
  \institution{\authorBaffil}
  \city{Logan}
  \country{USA}
}
\email{\authorBemail}

\author{\authorFname}
\affiliation{%
  \institution{\authorFaffil}
  \city{Logan}
  \country{USA}
}
\email{\authorFemail}

\author{\authorAname}
\affiliation{%
  \institution{\authorAaffil}
  \city{Logan}
  \country{USA}
}
\email{\authorAemail}
\authornote{Corresponding author}


\input{sections/0-abstract/abstract}

\maketitle

\input{sections/1-intro/introduction}

\input{sections/2-background/background}
\input{sections/3-design/0_overview}

\input{sections/4-implement/implementation}

\input{sections/5-evaluation/0_evaluation}

\input{sections/6-discussion/discussion}
\input{sections/7-threats/threats}
\input{sections/concl}

\bibliographystyle{ACM-Reference-Format}
\bibliography{reference}


\end{document}

%% file: common/preamble.tex
\usepackage[utf8]{inputenc}
\usepackage[english]{babel}

\usepackage{graphicx}
\DeclareGraphicsExtensions{.pdf,.png,.jpg}

\usepackage{booktabs}
\usepackage{multirow}
\usepackage{enumitem}
\usepackage{makecell}
\usepackage{relsize}

\usepackage{xcolor}
\usepackage[most]{tcolorbox}
\usepackage[normalem]{ulem}
\usepackage{colortbl}

\definecolor{darkgreen}{RGB}{0,100,0}
\definecolor{codegreen}{rgb}{0,0.5,0}
\definecolor{codepurple}{rgb}{0.58,0,0.82}
\definecolor{codegray}{rgb}{0.5,0.5,0.5}
\usepackage{listings}
\lstdefinestyle{mystyle}{
  commentstyle=\color{codegreen},
  keywordstyle=\bfseries,
  stringstyle=\color{codepurple},
  basicstyle=\ttfamily\scriptsize,
  breaklines=true,
  captionpos=b,
  keepspaces=true,
  tabsize=2
}
\usepackage[noend,ruled,linesnumbered]{algorithm2e}

\usepackage{tikz}
\usetikzlibrary{arrows.meta,positioning,shapes,calc} 

%% file: common/commands.tex
\newcommand{\find}[1]{%
\begin{tcolorbox}[tile,size=fbox,boxsep=2mm,boxrule=0pt,top=0pt,bottom=0pt,
borderline={0.6mm}{0pt}{black!66!white},colback=black!5!white]
\em #1
\end{tcolorbox}
}

\newcommand{\denseitems}{\setlength{\itemsep}{1pt}\setlength{\parskip}{0pt}}

\newboolean{COMMENTSON} 
\setboolean{COMMENTSON}{true}   
\ifthenelse{\boolean{COMMENTSON}}
{

}

\definecolor{DarkOrange}{rgb}{0.8,0.3,0.0} 
\definecolor{DarkCyan}{rgb}{0.0, 0.55, 0.55}
\definecolor{codegreen}{rgb}{0,0.6,0}
\definecolor{codegray}{rgb}{0.5,0.5,0.5}
\definecolor{codepurple}{rgb}{0.58,0,0.82}
\definecolor{backcolour}{rgb}{0.95,0.95,0.92}

\newcommand{\papertitle}{Beyond Source: An Empirical Study of Python Bytecode Security Risks}
\newcommand{\papertitleshort}{Beyond Source: An Empirical Study of Python Bytecode Security Risks}

\newcommand{\tech}{\mbox{\textsc{PycLens}}}

\newcommand{\paperkeywords}{Python bytecode, package security, software supply chain, program analysis, runtime robustness}

\newcommand{\authorAname}{Wen Li}
\newcommand{\authorAaffil}{Utah State University}
\newcommand{\authorAemail}{awen.li@usu.edu}

\newcommand{\authorBname}{Baihong Chen}
\newcommand{\authorBaffil}{Utah State University}
\newcommand{\authorBemail}{b.chen@usu.edu}

\newcommand{\authorFname}{Tian Xie}
\newcommand{\authorFaffil}{Utah State University}
\newcommand{\authorFemail}{tian.xie@usu.edu}


%% file: sections/0-abstract/abstract.tex

\begin{abstract}
Python package security is largely source-centric, 
yet Python runtimes can execute bytecode directly through \texttt{.pyc} files, 
compiled-only modules, and marshalled code objects, 
creating an inspection-execution gap. 
We present an empirical study of Python bytecode as a security artifact. 
We measure bytecode exposure in PyPI distributions, 
evaluate practical analyzability using version-aware tooling, 
assess CPython runtime robustness under adversarial bytecode, 
and test source-level reproduction of bytecode findings. 
Across 1{,}034{,}843 collected PyPI artifacts, 
we identify 7{,}388 bytecode-containing artifacts, 
including 228{,}578 \texttt{.pyc} files and 28{,}193 artifact-local source-less \texttt{.pyc} files. 
For modern CPython 3.8--3.14 bytecode,
at least one selected decompiler emits source for 204{,}901 of 204{,}904 in-scope files,
a result measuring emission rather than verified functional equivalence.
Tools are non-robust: observed PyPI bytecode triggers managed-code exceptions and timeouts, 
while adversarial mutated bytecode also drives decompilers into native process failures; 
together these outcomes yield 17 distinct robustness signatures.
Fuzzing produces 1,009 stack-deduplicated runtime findings dominated by
pointer-dereference symptoms; 261 groups exhibit potential memory-corruption
characteristics, and at least 91.7\% of groups reach execution beyond the
documented-unsafe ingestion boundary.
None reproduce from ordinary Python source. 
Bytecode is thus a visible ecosystem artifact, a practical analysis target, and a security-relevant 
interpreter input whose behavior need not match source-level behavior.
\end{abstract}

%% file: sections/1-intro/introduction.tex
\section{Introduction}\label{sec:intro}

Python package security is usually studied 
through source code, metadata, dependencies, and installation behavior. 
Source-centric analysis is necessary because these artifacts are visible 
to package scanners, dependency analyzers, malware detectors, and program-analysis pipelines. 
Prior studies have shown that package repositories such as PyPI are security-critical distribution channels for dependency risk, malicious packages, typosquatting, and supply-chain attacks~\cite{duan2021measuring,ohm2020backstabber,alfadel2023empirical,guo2023pypi,taylor2020spellbound,gao2025malguard}. 
These studies motivate analyzing the artifacts users actually install, 
but most security workflows still treat Python source and package metadata as the primary evidence.

That view is incomplete. 
Python runtimes execute code objects whose instruction streams are Python bytecode, 
and those code objects can be stored as \texttt{.pyc} files, 
embedded in compiled-only modules, loaded dynamically through marshalled payloads, 
or distributed together with ordinary source files. 
As a result, 
the representation executed by the interpreter 
can differ from the representation inspected by a source-centric security workflow. 
This difference is not merely an analysis inconvenience: 
bytecode can be consumed directly by Python runtimes, making untrusted or malformed bytecode part of the interpreter-facing attack surface.

Existing work leaves this bytecode surface only partially characterized. 
Bytecode tools can load, disassemble, 
or decompile many \texttt{.pyc} files~\cite{python-marshal,python-dis,decompylepp,pylingual}, 
and runtime fuzzing has exposed failures in language implementations~\cite{manes2018fuzzing,bohme2016coverage,li2023pyrtfuzz}. 
However, 
these lines of work do not establish how often bytecode appears in Python distributions, 
whether it lacks artifact-local source inside those artifacts, 
whether current tools can analyze the observed bytecode in version-aware settings, 
whether malformed bytecode can trigger runtime failures in modern interpreters, 
or whether bytecode-level findings correspond to ordinary source-level behavior.

This gap matters for three reasons. 
First, 
bytecode can be present in distribution artifacts even when nearby source is missing or incomplete, 
so a source-only inventory may miss executable content. 
Second, 
bytecode analysis is inherently version-sensitive: 
loading, disassembly, and decompilation depend on bytecode formats and tool support. 
Third, 
malformed bytecode can exercise interpreter states that ordinary source compilation does not produce, 
complicating how analysts should interpret bytecode-level crashes. 
More fundamentally, 
bytecode-level behavior and source-level behavior need not coincide. 
Even when bytecode can be inspected or partially decompiled, 
it remains unclear whether security-relevant bytecode behaviors 
can be reproduced through ordinary Python source. 
This distinction complicates both vulnerability triage and the interpretation of interpreter-level failures.

This paper studies Python bytecode as a first-class security artifact. 
We use the term \emph{security risk} broadly to include conditions that affect security-relevant visibility, availability, analysis integrity, and native-process safety when bytecode is distributed, inspected, or consumed across a trust boundary. We do not use the term to imply that every observed bytecode artifact is malicious or that every crash is an exploitable vulnerability.
We focus on PyPI distributions 
because they are the public package artifacts consumed 
by Python installation, dependency-resolution, and package-security workflows. 
The ecosystem component does not classify packages as malicious; it measures bytecode exposure, artifact-local source visibility, and tool behavior over observed PyPI artifacts. 
The study uses a staged empirical design that separates ecosystem transparency from adversarial interpreter robustness.
It first measures bytecode exposure and packaging patterns in collected PyPI artifacts. 
It then evaluates practical analyzability using version-aware CPython environments 
and selected bytecode tools. 
Next, 
it fuzzes version-matched CPython interpreters with mutated bytecode 
to study security-relevant robustness failures at the interpreter boundary. 
This fuzzing stage is complementary to the PyPI measurement: it does not claim that the observed PyPI bytecode corpus itself triggers the reported CPython crashes.
We use CPython as the concrete evaluation target 
because the observed \texttt{.pyc} artifacts, bytecode tags, standard loading and disassembly interfaces, 
and available interpreter builds are CPython-versioned; 
runtime findings are therefore scoped to CPython 3.8--3.14 
rather than all Python implementations. 
Finally, 
it evaluates whether selected source-recovery workflows 
can reproduce bytecode-level findings through ordinary Python source.

A central distinction throughout is the consumer of bytecode. 
Beyond the interpreter that executes it, bytecode is also processed as data by package scanners, 
malware-analysis pipelines, and decompilers whose purpose is to inspect potentially hostile code 
without running it. 
For these load-but-do-not-execute consumers, an attacker controls a data channel rather than a code 
channel, so robustness failures during bytecode parsing, loading, disassembly, or decompilation are 
security-relevant in their own right rather than redundant with code execution. 
We therefore separate two boundaries throughout the study: RQ3 evaluates the CPython bytecode execution 
boundary directly, while RQ2 and RQ4 show that non-executing bytecode-analysis consumers also experience 
data-channel robustness failures.
Table~\ref{tab:security-risk-taxonomy} maps these security-risk interpretations to the measurements that support them.

\begin{table}[t]
  \centering
  \caption{Security-risk interpretation of measured bytecode phenomena.}
  \label{tab:security-risk-taxonomy}
  \begin{tabular}{@{}p{0.23\linewidth}p{0.35\linewidth}p{0.32\linewidth}@{}}
    \toprule
    Risk class & Measured evidence & Interpretation boundary \\
    \midrule
    Package visibility risk
      & Artifact-local source-less \texttt{.pyc} files and dynamic-loading indicators
      & Executable content may be missed by source-only inspection; this does not imply maliciousness. \\
    Analysis anti-analysis risk
      & Tool timeouts, uncaught exceptions, and native process failures in bytecode-analysis tools
      & Bytecode can disrupt load-but-do-not-execute consumers; this does not imply code execution. \\
    Runtime hardening risk
      & CPython abnormal native terminations under mutated bytecode
      & Crash-reachable interpreter states identify hardening targets; this does not establish exploitability. \\
    Source-interpretation risk
      & Bytecode-level findings not reproduced through recovered source
      & Bytecode evidence should not be collapsed into source-level vulnerability claims. \\
    \bottomrule
  \end{tabular}
\end{table}

The study yields five main findings. 
First, 
Python bytecode is routinely encountered in PyPI distributions, 
including 28{,}193 artifact-local source-less \texttt{.pyc} files not directly visible through artifact-local source 
inspection; within the 21{,}541 in-scope CPython~3.8--3.14 artifact-local source-less \texttt{.pyc} files, roughly two-thirds are shipped import 
caches while one-third are bare compiled-only modules, the form most consistent with deliberate source 
omission. 
Second, 
modern CPython bytecode is overwhelmingly amenable to source emission with version-aware tooling: 
204{,}901 of 204{,}904 in-scope files reach L4, meaning that at least one selected decompiler emits source, though this level records 
successful decompiler emission rather than verified functional equivalence. 
Third, 
the analysis tools are themselves non-robust on bytecode inputs: observed PyPI bytecode triggers managed-code exceptions and timeouts, while adversarial mutated bytecode also drives decompilers into native process failures. Together, these outcomes yield 17 distinct tool-robustness signatures, exactly the data-channel failures the load-but-do-not-execute consumer faces. 
Fourth, 
separately from those analysis-tool failures, bytecode fuzzing across CPython 3.8--3.14 produces 1{,}009 stack-deduplicated runtime findings spanning 
multiple interpreter subsystems, dominated by pointer-dereference symptoms with a smaller 
memory-corruption subset (261 groups), and at least 91.7\% of these reach interpreter execution beyond 
the documented-unsafe ingestion boundary. 
Finally, 
none of these findings can be reproduced through the evaluated source-recovery workflows, 
revealing a measurable gap between bytecode-level and source-level behavior. 

We make the following \underline{contributions}:
\begin{itemize}
  \item We establish bytecode exposure as an empirical package-security problem rather than an anecdotal 
  artifact. Our PyPI-scale measurement shows where bytecode appears across distribution formats, how 
  often it lacks artifact-local source, which bytecode versions dominate the observed ecosystem, and how 
  the in-scope artifact-local source-less \texttt{.pyc} files divide between accidental cache leakage and deliberate compiled-only modules.

  \item We show that bytecode transparency cannot be reduced to source presence or a single decompiler 
  result. By separating loading, disassembly, decompilation, artifact-local source status, emission versus 
  faithfulness, and per-tool failures, the study provides a version-aware analyzability model for 
  bytecode-aware package triage, and shows that the analysis tools fail on real bytecode and 
  must be treated as part of the attack surface.

  \item We demonstrate that malformed bytecode exposes a distinct security-relevant CPython runtime-robustness boundary. 
  Version-matched bytecode fuzzing across CPython 3.8--3.14 produces stack-deduplicated crash groups 
  dominated by pointer-dereference failures across interpreter subsystems, the large majority reaching 
  execution beyond the documented-unsafe ingestion path, motivating bytecode validation and 
  runtime-hardening work beyond source-level testing. These findings are not automatically interpreted as exploitable vulnerabilities.

  \item We identify a source-reproduction gap for bytecode-level findings. The selected source-recovery 
  workflows do not reproduce any confirmed bytecode-fuzzing finding from ordinary Python source, showing 
  why bytecode-level evidence should be reported and triaged separately from source-level vulnerability 
  claims.
\end{itemize}

The central result is that bytecode 
introduces a measurable gap between what package-security workflows inspect, 
what Python runtimes execute, 
and what source-level artifacts can faithfully represent. 
Bytecode is visible in real package distributions, 
remains analyzable with modern tooling, can trigger interpreter-level runtime failures, 
and yet does not necessarily admit source-level reproduction. 
Security analyses that reason solely about source artifacts 
therefore risk overlooking behaviors that emerge at the bytecode layer. 
{The rest of the paper follows this staged empirical design.} 
Section~\ref{sec:background} gives the background, related work, and threat model. 
Section~\ref{sec:study} defines the study goals and methodology. 
Section~4 describes the empirical pipeline and {\tech} implementation. 
Section~\ref{sec:results} reports the four research questions, 
and Sections~\ref{sec:diss} and~\ref{sec:threats} discuss implications and threats to validity.

\vspace{3pt}
\noindent
\textbf{Open Science.}
To support transparency and reproducibility, we release the source code of {\tech}, 
together with the scripts, configurations, and datasets required to reproduce the study results, at \href{https://github.com/Cailbehumble/PycLens}{\underline{\tech}}. 
The artifact package includes support for dataset construction, 
bytecode analysis, runtime robustness testing, and source-reproduction evaluation.

%% file: sections/2-background/background.tex
\section{Background, Related Work and Threat Model}\label{sec:background}

Understanding the security implications of Python bytecode 
requires reasoning across three layers: 
package distributions, bytecode-analysis tooling, and interpreter execution. 
This section introduces the background needed to connect these layers. 
We first review Python bytecode and packaging mechanisms, 
then position the study within prior work on package security, bytecode analysis, and runtime testing. 
Finally, 
we define the threat model that motivates 
the empirical investigation of bytecode exposure, analyzability, runtime behavior, and source-level reproducibility.

\subsection{Background Knowledge}

\noindent
\textbf{Python Bytecode and Code Objects.}
Python source follows a compilation pipeline from source text 
to abstract syntax tree, code object, and bytecode execution.  
A code object contains the bytecode instruction stream together 
with constants, names, local-variable metadata, free and cell variables, flags, line and position information, and, in newer CPython versions, exception-table metadata~\cite{python-dis}.  
CPython serializes code objects using the \texttt{marshal} format~\cite{python-marshal} and stores import caches as \texttt{.pyc} files under pycache directories.  
These artifacts are version-sensitive: 
opcode sets, instruction layouts, stack effects, and metadata formats change across Python releases~\cite{python-dis,pep552}.
Although \texttt{.pyc} files are often described as caches, 
they are executable representations consumed by the interpreter.  
Python can import bytecode-only modules, load code objects from custom loaders~\cite{python-importlib}, 
execute unmarshalled code through \texttt{exec}, and ship compiled payloads inside application bundles.  
Therefore,
bytecode is not merely a performance optimization;
it is an executable artifact that forms part of both the attack surface and the analysis surface.
This matters even when a package also contains source, 
because the source file and the bytecode artifact may differ in version, path, timestamp, or content.

\vspace{3pt}
\noindent
\textbf{Python Distribution Artifacts and Bytecode Security.}
Python packages are commonly published as source distributions and wheel distributions, 
the standard distribution formats described 
by the Python packaging ecosystem~\cite{python-packaging-flow,python-package-formats}.  
A source distribution is a source-oriented archive, often a \texttt{.tar.gz} file, 
that contains the project files needed to build or install the package.  
A wheel is a built distribution, 
stored as a \texttt{.whl} archive, that contains files prepared for installation together with package metadata.  Although wheels are built distributions, 
they often still contain ordinary Python source files; 
the distinction is about the distribution format and installation workflow, not whether \texttt{.py} files are present. 
Under normal packaging guidance, wheel archives are not expected to contain generated \texttt{.pyc} cache files~\cite{python-package-formats, pep3147}; therefore, bytecode observed inside wheels is best interpreted as packaging-hygiene evidence, compiled-only shipping, or another nonstandard packaging pattern rather than ordinary wheel behavior.

This distinction matters because Python package distributions 
are a primary object of supply-chain security analysis.  
Prior studies use package-manager ecosystems, including PyPI, 
to measure supply-chain attacks, package vulnerabilities, and malicious-package detection~\cite{duan2021measuring,alfadel2023empirical,gao2025malguard}.  
These workflows inspect the artifacts that users and installers consume, 
so visibility is bounded by what the distribution file contains.  
We therefore treat wheels and source distributions as distinct artifact strata:
they are produced by different release workflows and may expose different executable content for the same package version.

Bytecode artifacts create a specific transparency problem inside these distributions.  
Package distributions may accidentally include cached bytecode; 
release systems may ship bytecode without source; 
bundlers may embed compiled modules; and malicious packages may hide payloads in marshalled code objects or compiled-only modules.  
Within either wheel or source distribution, 
we treat a bytecode file as source-present only when the corresponding \texttt{.py} file is present in the same artifact under the expected path.  
Otherwise, 
we classify it as an artifact-local source-less \texttt{.pyc} file, 
even if source may exist elsewhere in the project repository or in another distribution artifact.

The challenge is that source-level visibility is no longer guaranteed.  
A nearby \texttt{.py} file may be absent, stale, or unrelated to the \texttt{.pyc} 
that will be loaded.  
Decompilers may recover readable source for common cases, 
but they are not a complete substitute for bytecode analysis.  
A failed decompilation does not imply benign behavior, 
and a successful decompilation does not guarantee semantic equivalence after recompilation.  
For this reason, source presence, source matching, tool recoverability, 
and runtime behavior must be measured as separate properties.

\subsection{Related Work} \label{sec:related}

\noindent
\textbf{Python Package and Supply-Chain Security.}
Prior work studies package ecosystems as security-critical distribution channels.  
Empirical work on PyPI characterizes package metadata, releases, dependencies, imports, 
and ecosystem growth~\cite{bommarito2019pypi}.  
Supply-chain security studies measure malicious packages 
and attacks across interpreted-language package managers, including PyPI, npm, 
and RubyGems~\cite{duan2021measuring,ohm2020backstabber}.  
Other work studies Python-package vulnerabilities and malicious-package detection in PyPI~\cite{alfadel2023empirical,guo2023pypi,gao2025malguard}, 
while typosquatting defenses analyze how users can be led to install attacker-controlled packages with confusable names~\cite{taylor2020spellbound}.  
These efforts establish PyPI as an important empirical security population.  
Our study complements them by focusing on a different artifact property: 
whether distributed packages contain executable bytecode 
whose source visibility and tool analyzability must be measured directly.

\vspace{3pt}
\noindent
\textbf{Python Decompilation and Bytecode Analysis.}
Python bytecode tooling includes reference mechanisms for loading and disassembling code objects,
such as CPython \texttt{marshal} and \texttt{dis}~\cite{python-marshal,python-dis}, 
and decompilers such as Decompyle++ and PyLingual~\cite{decompylepp,pylingual}.  
Among these, Decompyle++ provides an independent native C++ implementation for Python bytecode decompilation; in this study, Decompyle++ denotes the evaluated implementation from the \texttt{zrsx/pycdc} repository, while PyLingual is oriented toward modern Python~3 decompilation.
Related tools also use bytecode decompilation 
to explain specific Python runtime systems; for example, \texttt{depyf} exposes bytecode 
generated by the PyTorch compiler by connecting in-memory code objects back 
to source-level views~\cite{you2024depyf}.
These tools demonstrate that bytecode can often be inspected or recovered,
but they do not establish how analyzable real-world bytecode is at ecosystem scale.
More broadly, 
decompiler studies in other bytecode ecosystems show that source recovery 
can fail syntactically or semantically and that different decompilers recover different subsets of inputs~\cite{harrand2019java}.  
This motivates our separation of loading, disassembly, decompilation, and source-reproduction outcomes.

\vspace{3pt}
\noindent
\textbf{Runtime Testing and Fuzzing.}
Fuzzing is a standard technique for exposing parser and runtime robustness failures~\cite{manes2018fuzzing,bohme2016coverage,fioraldi2020afl++}.  
Prior work has improved seed generation, scheduling, and feedback mechanisms for fuzzing programs and language runtimes~\cite{wang2017skyfire,yue2020ecofuzz,li2023polyfuzz}.  
Python-specific runtime fuzzing has studied bugs 
in Python runtimes through collaborative source/runtime-level fuzzing~\cite{li2023pyrtfuzz}.  
Source-based package-security analysis and source/runtime-level fuzzing 
remain essential because source files, metadata, dependencies, installation scripts, 
and runtime APIs expose many supply-chain and interpreter behaviors~\cite{duan2021measuring,alfadel2023empirical,guo2023pypi,li2023pyrtfuzz}.  
Our study differs in input level and interpretation: 
it mutates version-specific \texttt{.pyc} bytecode seeds 
and then tests whether bytecode-level findings are reproduced through selected source-recovery workflows.  
This separates interpreter behavior under malformed bytecode from ordinary source-level reachability.

\vspace{3pt}
\noindent
\textbf{Program Analysis Beyond Source.}
Security analysis often needs representations beyond source, 
including intermediate representations, bytecode, binaries, dynamic traces, 
and cross-language program representations.  
Compiler and analysis infrastructures such as LLVM and Soot 
demonstrate the value of lower-level representations for whole-program analysis~\cite{lattner2004llvm,lam2011soot}, while cross-language analysis work shows 
that security-relevant behavior may cross source-language boundaries~\cite{tan2007ilea,li2023polyfuzz}.  
The same lesson applies to Python bytecode: a source representation is useful but not always complete.  
A package may expose only a source-level loader 
while the behavior of interest resides in a marshalled code object, a compiled-only module, or a bytecode artifact that targets a different CPython version.  
Our study focuses on this gap between source-visible and executable Python code objects 
in real package distributions and interpreter inputs.

\subsection{Threat Model}\label{sec:threat}
The study distinguishes bytecode observed in ordinary package distributions 
from adversarial bytecode inputs.  
In ordinary packaging workflows, 
bytecode may appear as a build byproduct, cache file, or bundled artifact.  
Such bytecode is usually treated as part of a selected package 
once the package is installed or analyzed.  
We do not assume that every observed \texttt{.pyc} file is malicious; 
instead, these artifacts motivate measuring whether 
source-level inspection remains transparent.
The adversarial setting is different.  
An attacker may publish, replace, or tamper with bytecode 
so that a Python runtime receives inputs that were not produced 
by the normal source compiler path.  
We therefore consider two complementary risks: 
bytecode in public distributions may reduce transparency 
when source is missing or mismatched, and malformed bytecode 
may exercise interpreter behavior that ordinary Python source does not reach.

A natural objection is that any actor able to deliver malformed bytecode 
to an interpreter could instead deliver \emph{valid} bytecode that simply executes, 
so that crashing the runtime offers no capability the attacker does not already hold.  
This reasoning applies to the import path, where a planted \texttt{.pyc} 
is loaded \emph{as code}.  
It does not apply to a distinct class of consumers 
that process untrusted bytecode \emph{as data} without executing it: 
package scanners, malware-analysis pipelines, decompilers, 
and bytecode-inspection services whose explicit purpose 
is to inspect potentially hostile code safely.  
For these consumers the attacker controls a data channel rather than a code channel, 
so a robustness failure reachable through bytecode parsing, loading, or disassembly 
is not redundant with code execution.  
A crash or non-terminating hang yields an availability or anti-analysis primitive 
against the inspection pipeline, 
whereas a memory-safety failure on the loading or disassembly path 
constitutes a corruption primitive in a process 
the attacker could not otherwise influence.  
A related consumer is any service that deserializes marshalled code objects 
across a trust boundary; although consuming untrusted marshal data 
is explicitly {discouraged~\cite{python-marshal}}, 
that guidance itself confirms the consumer arises in practice.  
We therefore distinguish two threat settings. The first is the
interpreter-execution setting, where bytecode is delivered to a CPython
runtime and executed under a version-matched interpreter. This setting
captures the bytecode execution boundary exposed when malformed or
adversarial \texttt{.pyc} inputs reach the interpreter. The second is the
non-executing analysis setting, where package scanners, malware-analysis
pipelines, decompilers, and bytecode-inspection services process bytecode
as data rather than executing it. For these consumers, failures during
bytecode loading, parsing, disassembly, or decompilation are not redundant
with code execution: the attacker controls an input data channel rather
than an executed-code channel.

Runtime findings are not automatically treated as exploitable vulnerabilities; 
rather, they are evidence that bytecode execution constitutes 
a distinct interpreter boundary 
that should be evaluated separately from source-level behavior.  
Although alternative Python implementations exist, 
the observed bytecode tags, analysis tooling, and runtime experiments 
are all organized around CPython bytecode.  
Consequently, the study focuses on CPython 
rather than attempting to generalize findings 
across all Python implementations.  
All version-aware analyses, runtime experiments, 
and source-reproduction evaluations are therefore scoped 
to CPython bytecode formats and interpreter behavior.

%% file: sections/3-design/0_overview.tex
\section{Study Goals and Research Questions}\label{sec:study}

This study investigates whether Python bytecode should be treated as a first-class security artifact 
rather than merely as an incidental cache format. 
To answer this question, 
the study uses a staged empirical design that separates package-distribution evidence, tool analyzability, adversarial interpreter robustness, and source-level interpretation.
We first measure how bytecode appears in real PyPI distributions. 
We then evaluate whether the observed bytecode is practically analyzable using version-aware tooling. 
Next, 
we examine how CPython behaves when executing malformed bytecode inputs. 
Finally, 
we test whether bytecode-level findings can be reproduced 
through ordinary Python source recovered by selected source-recovery workflows.

The four goals separate ecosystem evidence, tool capability, interpreter behavior, and source-level interpretation. 
G1 characterizes bytecode exposure in distribution artifacts, including prevalence, source visibility, and packaging structure. 
G2 measures the practical analyzability achieved by selected bytecode-analysis tools. 
G3 evaluates security-relevant CPython runtime-robustness failures triggered by malformed or adversarial bytecode. 
G4 examines whether source-recovery workflows can bridge bytecode-level findings back to ordinary Python source. 
This separation ensures that each claim is tied to a specific input population, measurement procedure, and interpretation scope.

\begin{description}[style=unboxed,leftmargin=0cm]
\denseitems
\item[\textbf{RQ1 (G1):}] \emph{What packaging patterns characterize bytecode exposure in Python distributions?}
\item[\textbf{RQ2 (G2):}] \emph{What practical analyzability can existing bytecode tools achieve on observed Python bytecode?}
\item[\textbf{RQ3 (G3):}] \emph{What security-relevant robustness failures arise when CPython executes mutated bytecode?}
\item[\textbf{RQ4 (G4):}] \emph{Can selected source-recovery workflows reproduce bytecode-level findings through ordinary Python source?}
\end{description}

The ordering is deliberate and follows the two threat settings defined in
Section~\ref{sec:threat}. RQ1 first establishes package-distribution
exposure: how often bytecode appears in PyPI artifacts, how it is packaged,
and whether it lacks artifact-local source within the distributed artifact. This
motivates bytecode-aware inspection. RQ2 then evaluates the non-executing
analysis setting by measuring how effectively current tools can load,
disassemble, and decompile the observed modern CPython bytecode when
bytecode is processed as data. RQ3 shifts to the interpreter-execution
setting by executing mutated \texttt{.pyc} inputs under version-matched
CPython interpreters, thereby evaluating the runtime boundary exposed by
adversarial bytecode. Finally, RQ4 returns to source-level interpretation by
testing whether bytecode-level runtime findings can be translated through
source-recovery workflows into ordinary Python source that preserves the
observed behavior after recompilation and rerun. Together, the four questions separate ecosystem exposure, practical analyzability, adversarial interpreter robustness, and source-level interpretation. RQ3 and RQ4 are complementary robustness experiments and do not claim that the observed PyPI bytecode corpus itself triggers the reported CPython crashes.

\section{Methodology}\label{sec:method}

\begin{figure}[htp]
\centering
\scriptsize
\includegraphics[width=0.98\textwidth]{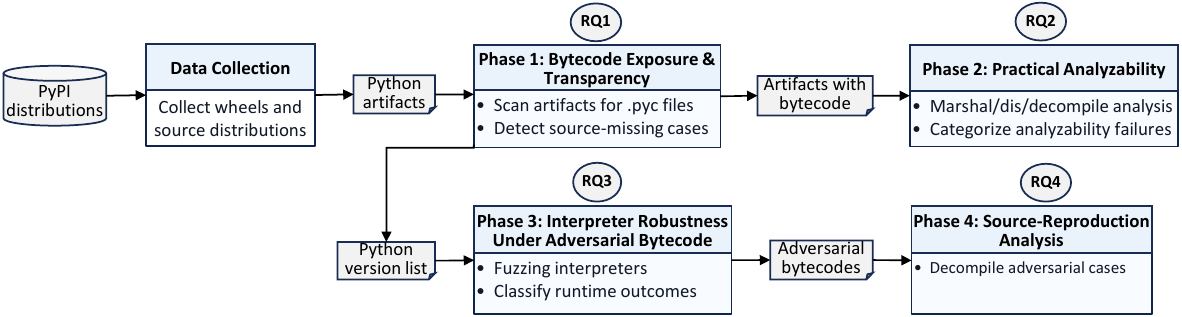}
\caption{Overview of the empirical study pipeline. The four phases correspond to bytecode exposure and packaging (RQ1), practical analyzability (RQ2), interpreter robustness (RQ3), and source reproduction (RQ4).}
\label{fig:pipeline-overview}
\end{figure}

This study investigates Python bytecode as a package-distribution artifact, 
an analysis target, a runtime input, and a source-recovery challenge. 
As illustrated in Figure~\ref{fig:pipeline-overview}, 
the methodology follows the same staged structure as the research questions.
The study first measures bytecode exposure in real PyPI distributions, 
then evaluates practical analyzability using version-aware tooling, 
next examines CPython runtime behavior under adversarial bytecode inputs, 
and finally tests whether bytecode-level findings can be reproduced through ordinary Python source.

The pipeline consists of four stages aligned with the research questions. 
First, 
the data-collection stage retrieves PyPI wheels and source distributions 
and inventories bytecode exposure, source visibility, packaging structure, dynamic-loading indicators, and observed bytecode versions. Second, 
the practical-analyzability stage evaluates whether selected tools can successfully load, disassemble, 
and decompile the observed bytecode artifacts. 
Third, 
the interpreter-robustness stage uses version-matched CPython environments 
to execute mutated bytecode and characterize security-relevant robustness failures. 
Fourth, 
the source-reproduction stage attempts 
to recover source from adversarial bytecode findings and determines whether source-derived bytecode 
can reproduce the original runtime behavior.

The remainder of this section follows the structure of Figure~\ref{fig:pipeline-overview}. 
We first describe data collection and study-population construction. 
We then present the methodology for measuring bytecode exposure and packaging characteristics (RQ1), 
practical bytecode analyzability (RQ2), interpreter robustness under adversarial bytecode (RQ3), 
and source-level reproduction of bytecode findings (RQ4). 
These stages separate ecosystem exposure, practical analyzability, adversarial interpreter robustness, and source-level interpretation. RQ3 and RQ4 do not claim that the observed PyPI bytecode corpus itself triggers the reported CPython crashes.

\subsection{Data Collection}\label{sec:method:data}

The data-collection stage defines the study population 
and establishes the denominator used throughout the empirical analysis. 
Because the study concerns bytecode exposure in distributed Python software, 
the collection methodology focuses on publicly distributed package artifacts rather than installed environments, repositories, or deployment-specific snapshots.
We use PyPI because it is the central public package ecosystem for Python 
and has been widely used in prior empirical 
and security studies of package vulnerabilities, dependency risk, and malicious-package detection~\cite{duan2021measuring,alfadel2023empirical,gao2025malguard}. 
PyPI provides the distribution artifacts consumed by package installers, 
dependency-resolution pipelines, and package-security workflows, 
making it an appropriate population for studying bytecode exposure in practice. 
In addition, 
PyPI exposes package names, release metadata, distribution files, hashes, 
and upload timestamps through documented index and metadata APIs, 
enabling a clear, reproducible, and auditable collection process~\cite{pypi-index-api,pypi-json-api}. 
Concretely, the implementation enumerates package names from the Simple/Index interface and obtains latest-release file records from PyPI's JSON metadata response; the resulting manifests preserve the exact file URLs, hashes, sizes, upload-derived selection decisions, local paths, and download status used in the study.

The collection boundary is wheels and source distributions, the standard package distribution formats described by Python packaging documentation~\cite{python-packaging-flow,python-package-formats}.  This boundary targets artifacts at the point where Python software is distributed, before installation rewrites files, creates cache directories, or resolves dependencies.  Installed environments, containers, application bundles, and repository snapshots are important deployment contexts, but they have different construction processes and denominators; including them would mix distribution prevalence with installation- or application-specific effects.
Table~\ref{tab:data-collection-design} summarizes the collection choices that define the dataset denominator.

\begin{table}[htp]
  \centering
  \caption{Data-collection choices and purpose.}
  \label{tab:data-collection-design}
  \begin{tabular}{p{0.32\linewidth}p{0.6\linewidth}}
    \toprule
    Choice & Purpose \\
    \midrule
    PyPI artifacts & Public distribution population for Python packages, with documented package indexes, metadata, file URLs, hashes, and upload times~\cite{pypi-docs,pypi-index-api,pypi-json-api}. \\
    Latest release & Measures the package version most likely to be retrieved by users and inspected by package-security workflows at collection time. \\
    One wheel and one source distribution & Covers the two standard distribution formats~\cite{python-package-formats} while keeping the denominator package-balanced rather than upload-balanced. \\
    Five-year upload window & Fixes the scope to contemporary package releases and modern CPython bytecode/tool support. \\
    Selection rule and CSV manifests & Makes artifact selection reproducible and preserves download status, hashes, paths, and failures for audit. \\
    \bottomrule
  \end{tabular}
\end{table}

The collector enumerates package names from the PyPI simple index, queries PyPI's JSON metadata, and selects artifacts from each package's latest release.  We use the latest release because it is the version most directly exposed to package installers and package-security triage at collection time; historical releases answer a different question about longitudinal evolution.  For each selected release, the collector keeps at most one wheel and one source distribution uploaded within the five-year study window.  When multiple files of the same kind are available, candidates are filtered by distribution kind and upload time, sorted deterministically by filename, and selected up to the per-kind limit.  This rule makes the dataset reproducible and prevents packages with many platform-specific wheels from outweighing packages with a single distribution file.
The five-year window is a scope choice rather than a claim that older artifacts are unimportant.  It fixes the denominator to contemporary packaging practice and modern CPython bytecode/tool support, which is the population needed for the later version-aware analyses.  If the latest release has no selected wheel or source distribution within the window, the package contributes no artifact to the collected dataset.  Source distributions may appear as \texttt{.tar.gz}, \texttt{.zip}, or related archive formats; given the dominance of tar-based sdists, we report them separately from zip/other.

The collection policy shapes the denominator.  Because the dataset is package-balanced rather than upload-balanced, projects with many platform-specific wheels do not dominate the measurement.  We therefore analyze wheels and source distributions as separate strata and report the combined total only as a descriptive summary.
For each selected artifact, 
the collector records package metadata, distribution type, artifact hashes, local paths, and collection status. 
The scanner then records file inventories, bytecode-version evidence, artifact-local source-less \texttt{.pyc} files, and dynamic-loading indicators. 
All collection and scanning outputs are preserved as structured CSV datasets 
to support auditability and reproducibility. 
Importantly, 
ambiguous or incomplete observations are retained in the raw records rather than normalized away, 
allowing later analyses to distinguish observed evidence from interpretation.

\input{sections/3-design/1_phase1}

\input{sections/3-design/2_phase2}
\input{sections/3-design/3_phase3}

%% file: sections/3-design/1_phase1.tex
\subsection{Phase 1: Bytecode Exposure and Packaging}\label{sec:method:extract}

Phase~1 measures how Python bytecode is exposed within collected distribution artifacts 
and constructs the ecosystem evidence used for RQ1. 
The primary unit of analysis is the distribution artifact, 
while file-level counts are retained to distinguish a small number of artifacts 
containing many bytecode files from a large number of artifacts containing only a few. 
The phase relies exclusively on static archive inspection: 
the scanner examines wheels, source distributions, zip archives, tar archives, individual files, 
and extracted directories without installing or executing package code. 
This boundary reflects the goal of RQ1, 
which is to characterize what package-security workflows can observe directly from distributed artifacts.
Table~\ref{tab:phase1-scan-fields} summarizes the information recorded during scanning.

\begin{table}[htp]
  \centering
  \caption{Phase~1 scan fields and purpose.}
  \label{tab:phase1-scan-fields}
  \begin{tabular}{p{0.3\linewidth}p{0.62\linewidth}}
    \toprule
    Field & Purpose \\
    \midrule
    Artifact type & Separate wheels, source distributions, and other archives. \\
    Source files & Measure whether bytecode is accompanied by visible Python source. \\
    \texttt{.pyc} files / pycache directories & Measure bytecode exposure at artifact and file levels. \\
    Artifact-local source-less \texttt{.pyc} files & Bytecode files that lack a plausible source counterpart in the same artifact under the path-based matching rule. \\
    Version evidence & Record magic numbers and filename tags for later version-aware analysis. \\
    Dynamic-loading indicators & Flag source that may load or construct bytecode at runtime. \\
    \bottomrule
  \end{tabular}
\end{table}

The recorded fields capture three complementary dimensions: exposure, source visibility, and packaging structure. 
Exposure is measured at both artifact and file levels because these denominators answer different security questions. 
Artifact-level prevalence estimates how frequently analysts encounter bytecode in distributed packages, 
whereas file-level prevalence estimates the amount of bytecode that must be inspected once such an artifact is identified. 
Source visibility is measured by whether a \texttt{.pyc} file 
has a plausible corresponding source file according to package layout conventions 
and Python's standard pycache naming scheme~\cite{python-py-compile}. 
Packaging structure is characterized through distribution type, 
pycache evidence, artifact-local source-less \texttt{.pyc} files, dynamic-loading indicators, and version information. 
These checks are intentionally conservative and artifact-local: 
they neither claim semantic equivalence between source and bytecode nor rule out the existence of source elsewhere.

The scanner preserves ambiguous observations 
rather than normalizing them away. Unknown bytecode versions, 
non-canonical filename tags, artifact-local source-less \texttt{.pyc} files, 
and dynamic-loading indicators remain explicit in the raw scan inventory, 
allowing later phases to apply their own selection criteria and interpretation rules. 
Dynamic-loading indicators are treated only as static evidence 
that code objects may be loaded or constructed at runtime. 
Operationally, the scanner searches source-like archive entries for a fixed set of textual patterns: \texttt{marshal.}, \texttt{marshal.loads(}, \texttt{importlib}, \texttt{SourcelessFileLoader}, \texttt{exec(}, \texttt{eval(}, and \texttt{types.CodeType}/\texttt{CodeType(}. 
These indicators do not demonstrate execution, validate intent, or identify the eventual runtime target. 
This separation ensures that RQ1 remains a descriptive measurement of ecosystem exposure 
while providing the version and packaging evidence required by the later analyzability and security-relevant runtime-robustness analyses.

%% file: sections/3-design/2_phase2.tex
\subsection{Phase 2: Practical Analyzability}\label{sec:method:source}
The second phase measures practical bytecode analyzability. 
RQ2 asks whether bytecode observed in PyPI distributions 
can be inspected using version-aware tools 
that are realistically available to analysts, 
rather than whether an ideal analysis system could recover every program. 
The unit of analysis is the individual \texttt{.pyc} file. 
For each bytecode-containing artifact, 
the phase extracts all \texttt{.pyc} entries, 
resolves their CPython version evidence, and evaluates 
whether selected tools can load, disassemble, or decompile them. 
The outcome is the highest analysis level achieved for each file.
We define five analyzability levels, ordered by the kind of analysis each supports, from basic 
code-object loading to source-level recovery:
\begin{itemize}
  \item \textbf{L0: Not loadable.} The bytecode cannot be loaded as a code object with the 
  version-matched runtime.
  \item \textbf{L1: Loadable only.} The code object can be loaded, but instruction-level disassembly is 
  not obtained.
  \item \textbf{L2: Disassemblable.} Bytecode instructions can be inspected, but no selected decompiler 
  recovers source.
  \item \textbf{L3: Partial decompilation.} At least one selected decompiler emits source for only part 
  of the bytecode input.
  \item \textbf{L4: Source emitted.} At least one selected decompiler emits source for the analyzed 
  bytecode input.
\end{itemize}

These levels measure observed tool recoverability, not semantic correctness. In particular, L4 denotes 
successful source \emph{emission}: at least one decompiler produced source for the input. It does not 
verify that the emitted source is functionally equivalent to the original bytecode, and we do not 
execute or otherwise check functional equivalence. A file counted at L4 may therefore, after 
recompilation, behave differently from the input. The level assignment thus measures whether tools can 
recover a source-level view, not whether that view faithfully reproduces the original program behavior; 
faithfulness is a distinct and stronger property that we do not claim for the analyzed corpus~\cite{harrand2019java}.
A decompile--recompile faithfulness check is applied in the source-reproduction phase 
(Section~\ref{sec:method:recovery}), but only to adversarial crash findings rather than to the L4 
ecosystem corpus.

We select tools according to five criteria that define the practical baseline for RQ2.
C1 (public availability),
C2 (modern Python compatibility),
C3 (practical adoption),
and C4 (maintenance or operational stability) ensure that the evaluated tools are accessible,
relevant to contemporary Python bytecode, and sufficiently reproducible for empirical evaluation.
C5 (methodological coverage) requires the selected set
to collectively cover the major stages of bytecode inspection:
code-object loading, instruction-level disassembly, and source-level decompilation.
Together,
these criteria define a representative measurement stack for practical analyzability rather than an exhaustive ranking of all Python bytecode-analysis tools.
Table~\ref{tab:tool-selection} summarizes the selected tools and the analysis stage each covers.
CPython \texttt{marshal} and \texttt{dis} provide reference loading and disassembly,
while Decompyle++ and PyLingual evaluate source-level recovery.
Together,
the selected tools satisfy all the selection criteria and span all analyzability levels.

\begin{table*}[t]
  \centering
  \caption{Selected bytecode-analysis tools and the analyzability stage each establishes. Pinned commits for the third-party decompilers are the revisions evaluated in RQ2 and RQ4; \texttt{marshal} and \texttt{dis} track the matched interpreter version rather than a fixed commit.}
  \label{tab:tool-selection}
  \small
  \begin{tabular}{@{}p{0.12\linewidth}p{0.32\linewidth}p{0.28\linewidth}p{0.22\linewidth}@{}}
    \toprule
    Tool & Functionality & Stage in analyzability pipeline & Pinned version / revision \\
    \midrule
    \texttt{marshal}~\cite{python-marshal}
      & Loads code objects from \texttt{.pyc}.
      & Code-object loading; reference baseline (L1).
      & Matched to the input bytecode version. \\
    \texttt{dis}~\cite{python-dis}
      & Disassembles bytecode instructions.
      & Instruction-level inspection (L2).
      & Matched to the input bytecode version. \\
    \texttt{Decompyle++}~\cite{decompylepp}
      & Native C++ cross-version decompiler.
      & Source recovery (L3--L4); independent cross-check.
      & \makecell[tl]{\texttt{zrsx/pycdc}\\ commit \texttt{175e0838e459}.} \\
    \texttt{PyLingual}~\cite{pylingual}
      & Modern CPython~3 decompiler.
      & Source recovery (L3--L4); primary path.
      & \makecell[tl]{Version \texttt{0.1.0}\\ commit \texttt{99c74eeff526}.} \\
    \bottomrule
  \end{tabular}
\end{table*}

The measurement is version-aware.
Python bytecode is version-sensitive and may change across CPython releases~\cite{python-dis}.
Therefore,
a \texttt{.pyc} file is included in tool analysis only when the Phase~1 scan
identifies a supported modern CPython tag and the corresponding runtime can be prepared.
In the current scope,
this means CPython~3.8 and later, 
the modern CPython subset that dominates the observed version evidence
and for which current interpreters and decompilers can be evaluated consistently.
CPython~3.7 and earlier, PyPy tags, unknown tags,
and noncanonical tags without an interpreter mapping remain in the raw scan inventory
but are excluded from tool-failure counts.
This prevents missing interpreters or unsupported runtimes from being misreported as tool limitations.

To make this version-aware analysis possible,
the phase prepares analysis runtimes before running bytecode tools.
It reads the CPython-version summary exported by Phase~1,
uses the recorded interpreter mapping to identify canonical CPython runtimes,
locates or installs matching interpreters when the host supports it,
and creates per-version tool environments for tools that depend on bytecode format.
Each analysis step records both status and failure reason,
including marshal failure, disassembly failure,
unsupported bytecode version, external-tool timeout, nonzero exit, and unavailable decompiler output.
These failure reasons are necessary for interpretation:
an environment miss, a loader failure, and a decompiler failure imply different limits.
The analysis proceeds in increasing order of abstraction.
First,
the version-matched runtime attempts to load the bytecode payload as a code object.
Second,
CPython \texttt{dis} attempts to produce an instruction-level representation.
Third,
the selected decompilers attempt source recovery.
The final analyzability level is assigned according to the strongest successful outcome,
while per-tool statuses and failure reasons are retained to explain why a file does not reach a higher level.

%% file: sections/3-design/3_phase3.tex
\subsection{Phase 3: Interpreter Robustness Under Adversarial Bytecode}\label{sec:method:runtime}

\begin{figure}[t]
  \centering
  \scriptsize
  \includegraphics[width=0.98\textwidth]{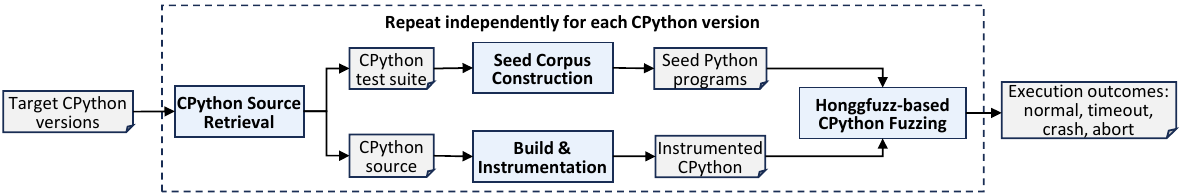}
  \caption{Version-aware workflow for evaluating CPython robustness under adversarial bytecode. 
  Each CPython version is processed independently using version-matched source retrieval, seed generation, interpreter instrumentation, and bytecode fuzzing.}
  \label{fig:rq3-fuzzing-workflow}
\end{figure}

The third phase evaluates CPython runtime robustness 
under adversarial bytecode inputs. 
RQ3 asks whether malformed bytecode is rejected through controlled Python-level failure mechanisms 
or whether it can drive the interpreter 
into abnormal termination states such as crashes, aborts, or persistent hangs.
Figure~\ref{fig:rq3-fuzzing-workflow} summarizes the version-aware workflow.  
The input is the set of target CPython versions observed in Phase~1 
for which a matching instrumented build can be prepared; 
unsupported legacy or non-CPython versions are recorded as out of scope for this phase.  
Each version is processed independently 
so that bytecode format, seed corpus, interpreter build, 
and fuzzing outcomes remain attributable to a specific CPython release.

The workflow begins by retrieving the CPython source for each version.  
The source tree provides both inputs for version-aware fuzzing: 
the unittest programs used as seed material and the code base used to build the fuzzing target.  
We use CPython's own unittest suite because it is maintained with the corresponding interpreter 
and exercises core language and runtime functionality~\cite{cpython-test-suite,cpython-devguide-tests}.  
The seed programs therefore provide version-correct and interpreter-maintained examples of valid bytecode generation. 
Starting from these seeds allows mutation to explore malformed bytecode states 
while avoiding confounding effects caused by version mismatches or manually constructed inputs.
Seed construction is bytecode-based.  
The extracted unittest programs are not fuzzed as source text; 
they are compiled with the matching CPython version into valid bytecode seeds 
for that release using Python's standard source-to-bytecode compilation interface~\cite{python-py-compile}.  
This preserves version correctness while ensuring that mutation starts from bytecode that CPython can normally produce. 
The fuzzer can then explore malformed bytecode states outside the source-compiler path 
without conflating those states with bytecode/interpreter mismatch. 
This design exercises byte-level malformation of serialized \texttt{.pyc}/marshal content; it does not model an attacker who carefully crafts a valid marshal stream or valid code object that imports normally, which remains a separate future-work boundary.

This bytecode-level input is what distinguishes the design from existing CPython fuzzing, 
which targets a different surface rather than a smaller one. Continuous efforts such as OSS-Fuzz and 
prior Python runtime fuzzing, including the two-level source/runtime fuzzing of 
Pyrtfuzz~\cite{ossfuzz,li2023pyrtfuzz}, mutate Python source, runtime API calls, or C-level inputs, and therefore 
reach the interpreter only through the ordinary source-to-bytecode compilation path. By mutating 
serialized bytecode seeds directly, our campaigns exercise the bytecode-ingestion and execution 
boundary that those campaigns do not target by construction. The goal is not to produce more crashes 
than source-level fuzzers, but to evaluate whether this distinct input surface is crash-reachable; 
because mutated bytecode can encode code-object states the source compiler never emits, 
Phase 4 
subsequently tests whether any resulting findings are reproducible through ordinary Python source~\cite{klees2018evaluating}.

For the execution target, the phase builds an instrumented CPython 
from the same source tree and stores the build separately for each version.  
Instrumentation is used to guide mutation toward new bytecode-processing states, 
following the standard coverage-guided fuzzing model~\cite{bohme2016coverage}; 
coverage is not treated as the study result.  
The goal is to exercise the bytecode-processing paths of the corresponding interpreter, 
not to compare source-level program behavior across releases.  
Keeping the source release, seed corpus, 
and interpreter build version-specific makes runtime outcomes attributable to the tested CPython release.

Finally, 
honggfuzz receives the version-specific bytecode seed corpus and the matching instrumented CPython build. 
We use honggfuzz because it is a local, coverage-guided mutation fuzzer 
for native programs and provides built-in crash and timeout collection~\cite{honggfuzz}. 
These properties align with the goal of RQ3: 
mutate bytecode inputs, execute them under bounded conditions, and preserve abnormal outcomes for analysis. 
The phase records normal execution, timeout, crash, abort, and other abnormal termination events. 
Bounded execution prevents nonterminating inputs from stalling campaigns 
and allows persistent hangs to be treated as a distinct outcome class. 
Crashes, aborts, and persistent timeouts are considered security-relevant robustness findings because they indicate failures at the bytecode execution boundary, although they are not automatically interpreted as exploitable vulnerabilities.

\subsection{Phase 4: Source-Reproduction Analysis}\label{sec:method:recovery}

The fourth phase evaluates whether bytecode-level findings can be reproduced through ordinary Python source. 
RQ4 does not ask whether a source-level reproducer exists in principle. 
Instead, it measures whether selected, practical source-recovery workflows 
can bridge the gap between a bytecode-level runtime finding and an equivalent source-level reproducer. 
The unit of analysis is a confirmed abnormal bytecode input produced by Phase~3, 
comprising the crashes and aborts confirmed in that phase.

Each finding is assigned to one of four mutually exclusive outcome categories 
based on the source-recovery stage it reaches. 
Source recovery may fail to emit source (\emph{no source}), 
produce source that fails to compile (\emph{compile failure}), 
or produce source-derived bytecode that does not reproduce the original finding on rerun 
(\emph{source rerun not reproduced}). 
A finding is classified as \emph{source reproduced} only if recovered source compiles 
under the matching CPython version~\cite{python-py-compile} 
and the resulting bytecode reaches the same behavior class as the original input. 
The inputs are the stack-deduplicated findings already confirmed abnormal in Phase~3, 
so confirmation is established upstream and is not repeated as a separate category here.

\begin{algorithm}[htp]
  \DontPrintSemicolon
  \KwIn{Confirmed abnormal bytecode input $b$ with Phase~3 behavior class $c_b$, CPython version $v$, selected recovery tools $T$}
  \KwOut{Source-reproduction category and per-tool failure stage}
  \ForEach{$t \in T$}{
    $s \leftarrow \textnormal{\textsc{RecoverSource}}(t,b,v)$\nllabel{alg:p4:recover}\;
    \If{$s = \textnormal{\textsc{Fail}}$}{
      $\textnormal{\textsc{Record}}(t,\textnormal{\textsc{NoSource}})$\;
      \textbf{continue}\;
    }
    $b_s \leftarrow \textnormal{\textsc{CompileSource}}(v,s)$\nllabel{alg:p4:compile}\;
    \If{$b_s = \textnormal{\textsc{Fail}}$}{
      $\textnormal{\textsc{Record}}(t,\textnormal{\textsc{CompileFailure}})$\;
      \textbf{continue}\;
    }
    $c_s \leftarrow \textnormal{\textsc{RunHarness}}(v,b_s)$\nllabel{alg:p4:rerun}\;
    \If{$\textnormal{\textsc{SameClass}}(c_s,c_b)$}{
      \Return \textnormal{\textsc{SourceReproduced}}$(t)$\nllabel{alg:p4:reproduced}\;
    }
    $\textnormal{\textsc{Record}}(t,\textnormal{\textsc{SourceRerunNotReproduced}})$\nllabel{alg:p4:diverge}\;
  }
  \Return \textnormal{\textsc{NotReproducedBySelectedTools}}\nllabel{alg:p4:notreproduced}\;
  \caption{Source-reproduction analysis for confirmed abnormal bytecode findings.}
  \label{alg:source-reproduction}
\end{algorithm}

Algorithm~\ref{alg:source-reproduction} implements the classification procedure. 
For each selected recovery tool, 
lines~\ref{alg:p4:recover}--\ref{alg:p4:compile} test whether the tool emits ordinary Python source 
and whether that source is valid for the matching interpreter. 
Lines~\ref{alg:p4:rerun}--\ref{alg:p4:diverge} rerun the source-derived bytecode 
and compare behavior classes rather than exact process details, 
so crashes and aborts are compared at the level relevant to RQ4. 
A key invariant of the procedure is version consistency: 
the original bytecode, its recorded behavior class, the recovered source, the recompiled bytecode, 
and the execution harness all use the same CPython version. 
This prevents reproduction outcomes from being confounded by interpreter-version differences. 

This design records the stage at which reproduction fails 
because different failure modes imply different interpretation limits. 
Source-recovery failure, source-validity failure, and source-rerun non-reproduction represent distinct 
obstacles and should not be collapsed into a single decompilation outcome~\cite{harrand2019java}. 
The distinction is particularly important because bytecode mutation 
can create code-object states that ordinary Python source may never generate. 
Consequently, a bytecode-level runtime finding should not automatically be interpreted as source-level 
reachable behavior. 
The decompile--compile--rerun workflow therefore serves as a bridge-validation step: 
it evaluates whether selected source-recovery workflows 
can preserve the behavior responsible for the original finding 
while keeping negative results explicitly tool-bounded 
rather than treating them as proof that no source-level reproducer exists.

%% file: sections/4-implement/implementation.tex
\subsection{Tool Implementation}\label{sec:method:tools}

{\tech} is implemented as a Python package 
that automates the staged empirical pipeline while preserving all intermediate observations as explicit artifacts. 
Its command-line interface orchestrates dataset construction, 
bytecode inspection, CPython-environment preparation, 
interpreter robustness testing, and source-reproduction analysis. 
Rather than exchanging hidden in-memory state, 
components communicate through structured artifacts stored in a shared workspace. 
This design supports reproducibility, auditability, and efficient reruns of long-running experiments.

\begin{figure}[htp]
  \centering
  \includegraphics[width=0.9\textwidth]{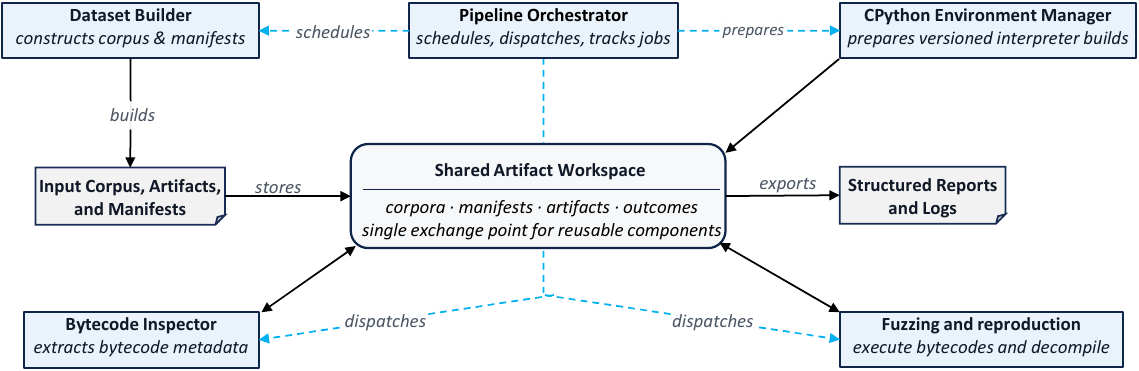}
 \caption{Tool-support architecture of {\tech}. A pipeline orchestrator coordinates dataset construction, bytecode inspection, CPython-environment preparation, and fuzzing/reproduction workflows through a shared artifact workspace that stores intermediate datasets, findings, reports, and logs.}
  \label{fig:tool-architecture}
\end{figure}

Figure~\ref{fig:tool-architecture} summarizes the tool-support architecture. 
At the center is a shared artifact workspace 
that stores all intermediate datasets, manifests, findings, reports, 
and logs exchanged between components. 
The pipeline orchestrator coordinates execution and dispatches tasks 
across the study workflow while tracking job status and dependencies.
The dataset builder retrieves PyPI metadata and distribution artifacts 
according to the collection policy and records acquisition outcomes in reproducible manifests. 
The bytecode inspector analyzes collected artifacts 
without importing package modules and records file inventories, 
bytecode-version evidence, source visibility, dynamic-loading indicators, and scan outcomes. 
The CPython environment manager prepares version-specific interpreter builds 
and tool environments required by the later phases.
The fuzzing and reproduction component supports 
both runtime robustness testing and source-reproduction analysis. 
It manages version-specific seed generation, instrumented CPython builds, 
fuzzing campaigns, crash collection, source recovery, recompilation, and behavioral-equivalence evaluation. 
All outputs are written back to the shared workspace and exported 
as structured reports and logs for subsequent analysis.

Across all stages, 
structured artifacts stored in the shared workspace 
preserve the intermediate evidence produced by each empirical stage. 
The workspace serves as the exchange point between components 
and the source for exported reports and logs. 
This design allows individual analyses to be rerun, audited, 
or extended without repeating expensive collection, decompilation, or fuzzing campaigns.

%% file: sections/5-evaluation/0_evaluation.tex
\section{Results}\label{sec:results}

This section reports the study results for the four research questions. 
Consistent with the study design in Section~\ref{sec:study} and the methodology in Section~\ref{sec:method}, 
RQ1 examines bytecode exposure and packaging structure,
RQ2 evaluates practical analyzability, 
RQ3 evaluates interpreter robustness under adversarial bytecode, 
and RQ4 evaluates source reproduction.

\noindent
{\bf Experiment Environment.}
All experiments are conducted in isolated Ubuntu 22.04 LTS environments on a machine 
equipped with a 13th Gen Intel Core i9-13900F processor (32 logical CPUs), 64 GB RAM, and 8 TB disk storage. 
Package collection retrieves public distribution artifacts directly from PyPI. 
Tool-based analyses use version-aware environments: 
CPython loading and disassembly are performed with the matching interpreter, 
Decompyle++ is executed as an independent native decompiler from \texttt{zrsx/pycdc} commit \texttt{175e0838e459}, 
and PyLingual is executed from a dedicated Python~3.12 environment using version \texttt{0.1.0} at commit \texttt{99c74eeff526}. 
Runtime robustness testing uses a common honggfuzz~\texttt{2.6} backend from source snapshot \texttt{01713bdad730}, 
while each evaluated CPython release is built separately as an instrumented fuzzing target. 
Unless otherwise specified, 
all experiments are executed under the same hardware and software configuration, 
and all fuzzing campaigns use identical resource allocations and time budgets. 
Abnormal fuzzing outcomes are treated as security-relevant bytecode-level findings 
that require manual triage and validation rather than being automatically classified as vulnerabilities.

\input{sections/5-evaluation/1_RQ1}
\input{sections/5-evaluation/2_RQ2}

\input{sections/5-evaluation/3_RQ3}

\input{sections/5-evaluation/4_RQ4}

%% file: sections/5-evaluation/1_RQ1.tex
\subsection{RQ1: Bytecode Exposure and Packaging}\label{sec:effmeasure}

RQ1 measures how Python bytecode artifacts are exposed and packaged in collected PyPI distributions.  
The analysis has three dimensions: 
artifact-level exposure, source visibility, and archive-level packaging structure.  
For each package, 
the collector keeps at most one wheel and one source distribution when available.  
This package-balanced design prevents projects with many uploaded files from dominating the measurement.  
It also makes the wheel and source-distribution row sizes similar by construction; 
they are separate packaging strata, not duplicate copies of the same artifact.  
We keep both rows because packaging format is part of the analysis, 
and we report the combined row only as a descriptive total over collected artifacts.  
We count artifacts containing \texttt{.pyc} files, pycache directories, artifact-local source-less \texttt{.pyc} files, dynamic-loading indicators, and observed CPython bytecode tags.  
Results are reported both per distribution artifact and per bytecode file because one artifact can contain many bytecode files.

\begin{table}[htp]
  \centering
  \caption{Bytecode exposure and source visibility by package-balanced PyPI artifact stratum. Wheel and source-distribution rows are the primary comparisons because the collection keeps at most one of each per package; their near-equal row sizes are therefore expected. The total row summarizes collected artifacts and is not used as a natural PyPI upload-frequency estimate. Percentages for artifacts with bytecode and dynamic-loading artifacts are computed over the number of artifacts in the row; percentages for artifact-local source-less \texttt{.pyc} files are computed over the number of \texttt{.pyc} files in the row.}
  \label{tab:prevalence}
  \begin{tabular*}{\textwidth}{@{\extracolsep{\fill}}lrrrrr@{}}
    \toprule
    Artifact stratum & \makecell{Number of\\artifacts} & \makecell{Artifacts with\\bytecode} & PYC file count & \makecell{Artifact-local\\source-less PYC files} & \makecell{Dynamic-loading\\artifacts} \\
    \midrule
    Wheel & 517{,}076 & 2{,}646 (0.51\%) & 76{,}058 & 8{,}818 (11.59\%) & 114{,}043 (22.06\%) \\
    Source distribution(tar) & 517{,}077 & 4{,}702 (0.91\%) & 151{,}472 & 19{,}301 (12.74\%) & 128{,}407 (24.83\%) \\
    Zip/other & 690 & 40 (5.80\%) & 1{,}048 & 74 (7.06\%) & 136 (19.71\%) \\
    \midrule
    All collected artifacts & 1{,}034{,}843 & 7{,}388 (0.71\%) & 228{,}578 & 28{,}193 (12.33\%) & 242{,}586 (23.44\%) \\
    \bottomrule
  \end{tabular*}
\end{table}

\subsubsection{Overview of Bytecode Exposure}
Table~\ref{tab:prevalence} shows that Python bytecode appears throughout the PyPI ecosystem. 
Across the collected artifacts, 7{,}388 artifacts contain at least one \texttt{.pyc} file, yielding a 
total of 228{,}578 bytecode files. Bytecode appears in both wheels and source distributions. 
The wheel result is particularly informative as packaging-hygiene evidence: because wheels are not 
expected to contain generated \texttt{.pyc} cache files under normal packaging guidance, the 76{,}058 
wheel-contained \texttt{.pyc} files indicate bytecode leakage, compiled-only shipping, or other 
nonstandard release patterns rather than ordinary wheel behavior. Although the artifact-level 
prevalence is below 1\% in both strata, the absolute volume of observed bytecode demonstrates that 
package-security workflows encounter bytecode at ecosystem scale.
The same table reports two further security-relevant dimensions. First, 28{,}193 of the 228{,}578 
\texttt{.pyc} files (12.33\%) are artifact-local source-less \texttt{.pyc} files under our conservative artifact-local source check: they 
lack a corresponding \texttt{.py} file within the same distribution artifact and are therefore not 
directly transparent to source-level inspection. Artifact-local source-less \texttt{.pyc} files do not necessarily indicate 
malicious intent, but they represent executable package content unavailable to artifact-local source 
inspection, and we characterize this subset in detail in Section~\ref{sec:sourceless-char}. 
Second, dynamic-loading indicators appear in 242{,}586 artifacts (23.44\%), including many that also 
contain bytecode, suggesting that bytecode frequently appears in packaging contexts where execution 
behavior cannot be inferred from static source files alone.

\subsubsection{Version Diversity}
\begin{table}[htp]
  \centering
  \caption{Bytecode-version distribution for all 228{,}578 observed \texttt{.pyc} files. The table uses the same version criterion as RQ2, so the CPython 3.8--3.14 counts are directly comparable with Table~\ref{tab:tool-analyzability}. Gray rows are outside the follow-up version-aware analyses; darker gray rows are non-CPython or unresolved bytecode.}
  \label{tab:bytecode-versions}
  \begin{tabular}{lrr}
    \toprule
    Resolved bytecode version & PYC file count & Percentage \\
    \midrule
    CPython 3.10 & 51{,}487 & 22.52\% \\
    CPython 3.11 & 38{,}320 & 16.76\% \\
    CPython 3.12 & 36{,}637 & 16.03\% \\
    CPython 3.8 & 31{,}633 & 13.84\% \\
    CPython 3.9 & 30{,}321 & 13.27\% \\
    CPython 3.13 & 13{,}274 & 5.81\% \\
    \textcolor{gray}{CPython 3.7} & \textcolor{gray}{9{,}954} & \textcolor{gray}{4.35\%} \\
    \textcolor{black!60}{Unknown} & \textcolor{black!60}{6{,}550} & \textcolor{black!60}{2.87\%} \\
    \textcolor{gray}{CPython 3.6} & \textcolor{gray}{5{,}789} & \textcolor{gray}{2.53\%} \\
    CPython 3.14 & 3{,}232 & 1.41\% \\
    \textcolor{gray}{CPython 3.5} & \textcolor{gray}{982} & \textcolor{gray}{0.43\%} \\
    \textcolor{gray}{CPython 3.4} & \textcolor{gray}{161} & \textcolor{gray}{0.07\%} \\
    \textcolor{gray}{CPython 2.7} & \textcolor{gray}{173} & \textcolor{gray}{0.08\%} \\
    \textcolor{gray}{CPython 2.6} & \textcolor{gray}{46} & \textcolor{gray}{0.02\%} \\
    \textcolor{gray}{CPython 3.3} & \textcolor{gray}{13} & \textcolor{gray}{0.01\%} \\
    \textcolor{gray}{CPython 3.1} & \textcolor{gray}{3} & \textcolor{gray}{0.00\%} \\
    \textcolor{gray}{CPython 3.2} & \textcolor{gray}{2} & \textcolor{gray}{0.00\%} \\
    \textcolor{gray}{CPython 3.15} & \textcolor{gray}{1} & \textcolor{gray}{0.00\%} \\
    \bottomrule
  \end{tabular}
\end{table}
Table~\ref{tab:bytecode-versions} shows that bytecode spans a wide range of Python versions. 
For consistency, Table~\ref{tab:bytecode-versions} and Table~\ref{tab:tool-analyzability} use the same 
version criterion. CPython 3.8--3.14 accounts for 204{,}904 files (89.64\%), which is the RQ2 analysis 
scope, while older CPython versions and unresolved tags account for the remaining 23{,}674 files. 
The dominance of multiple modern CPython generations, rather than a single version, is what makes 
version-aware loading and decompilation necessary; this diversity directly motivates the version-aware 
analysis pipeline evaluated in RQ2.

\subsubsection{Characterizing Artifact-Local Source-Less \texttt{.pyc} Files}\label{sec:sourceless-char}
Artifact-local source-less \texttt{.pyc} files are the subset most relevant to source-level transparency, since they carry 
executable code with no artifact-local \texttt{.py} counterpart. We therefore examine it more closely 
along two axes: how it is packaged, and what functionality it concerns.

\vspace{3pt}
\noindent
\textbf{Packaging structure.}
To distinguish accidental exposure from deliberate compiled-only distribution, we characterize the 
artifact-local source-less \texttt{.pyc} files by packaging structure. We restrict this characterization to the 21{,}541 in-scope CPython~3.8--3.14 artifact-local source-less \texttt{.pyc} files, the same modern-CPython scope used in 
RQ2, so that the population is consistent across research questions. Table~\ref{tab:sourceless} 
summarizes the result. By layout, 14{,}186 of these files (65.9\%) reside inside \texttt{\_\_pycache\_\_} 
directories, indicating import caches shipped with the distribution rather than compiled-only modules; 
they arise from accidental import caches, bundled environment caches, and shipped test caches. The 
remaining 7{,}355 files (34.1\%) are bare \texttt{.pyc} modules outside any cache directory: 
compiled-only modules (6{,}455), bundled application or environment payloads (538), and generated build 
artifacts (362). The two groups carry different security implications. Cache leakage is primarily a 
packaging-hygiene and transparency issue, whereas bare compiled-only modules are the form expected from 
deliberate source omission or bytecode obfuscation and are the subset an analyst should prioritize. By 
distribution type, these artifact-local source-less \texttt{.pyc} files appear predominantly in source distributions (13{,}636) and 
wheels (7{,}856); its presence in source distributions is notable because those archives are nominally 
source-oriented.

\begin{table}[htp]
  \centering
  \caption{Packaging characterization of the 21{,}541 in-scope CPython~3.8--3.14 artifact-local source-less \texttt{.pyc} files. The \emph{Layout} and \emph{Plausible cause} groups each independently partition the 21{,}541 files; cause is inferred from archive path shape. The \emph{Distribution type} group reports the same files by containing archive.}
  \label{tab:sourceless}
  \begin{tabular}{lr}
    \toprule
    Characterization & Files \\
    \midrule
    \multicolumn{2}{l}{\textit{Layout}} \\
    \quad Inside \texttt{\_\_pycache\_\_} (shipped import cache) & 14{,}186 \\
    \quad Bare \texttt{.pyc} module (no cache directory) & 7{,}355 \\
    \midrule
    \multicolumn{2}{l}{\textit{Plausible cause (from path shape)}} \\
    \quad Accidental import cache & 7{,}925 \\
    \quad Bare compiled-only module & 6{,}455 \\
    \quad Bundled environment cache & 5{,}208 \\
    \quad Accidental test cache & 1{,}053 \\
    \quad Bundled application/environment & 538 \\
    \quad Generated/build artifact & 362 \\
    \midrule
    \multicolumn{2}{l}{\textit{Distribution type}} \\
    \quad Source distribution (\texttt{tar}) & 13{,}636 \\
    \quad Wheel & 7{,}856 \\
    \quad Zip/other & 49 \\
    \bottomrule
  \end{tabular}
\end{table}

\vspace{3pt}
\noindent
\textbf{Topical exposure.}
Packaging structure indicates how artifact-local source-less \texttt{.pyc} files are shipped but not what functionality 
it concerns. To approximate which artifacts containing artifact-local source-less \texttt{.pyc} files touch sensitive functionality, we tag each 
artifact containing artifact-local source-less \texttt{.pyc} files by topic keywords matched against package and module identifiers. The categories 
are not mutually exclusive, and a match indicates topical relevance rather than suspicious behavior. 
Sixty-four artifacts containing artifact-local source-less \texttt{.pyc} files match the security-related category (for example, identifiers 
associated with cryptography, authentication, or access control); among these same artifacts, 25 also 
match network, 16 data, 16 testing, 9 CLI, 8 build, 8 machine-learning, and 5 plugin categories. These 
tags identify where artifact-local source-less \texttt{.pyc} files coincide with security-relevant functionality, and therefore 
where bytecode-aware inspection is most warranted; they are a static topical filter and do not by 
themselves establish benign or malicious behavior. We characterize packaging and topical exposure here 
and leave behavioral analysis of the decompiled artifact-local source-less \texttt{.pyc} set, including dynamic-loading, network, 
and obfuscation patterns, to future work, as noted in Section~\ref{sec:threats}.

\vspace{4pt}
Taken together, these findings establish that bytecode is a practical ecosystem artifact rather than a 
theoretical edge case. Bytecode appears across multiple packaging formats and CPython generations, 
frequently coexists with dynamic-loading mechanisms, and is not always accompanied by artifact-local source. 
Consequently, package-security workflows require bytecode-aware inventory and analysis rather than 
assuming that source files provide a complete view of executable package content.

\find{\textbf{Answer to RQ1.} 
Python bytecode appears across PyPI distribution artifacts, 
spans multiple packaging formats and CPython versions, and is not always accompanied by artifact-local source. 
The presence of 28{,}193 artifact-local source-less \texttt{.pyc} files 
and the frequent coexistence of bytecode with dynamic-loading mechanisms 
indicate that package-security workflows require explicit bytecode-aware inventory 
and analysis rather than relying solely on source files. 
Within the 21{,}541 in-scope CPython~3.8--3.14 artifact-local source-less \texttt{.pyc} files, roughly two-thirds are shipped import caches while 
one-third are bare compiled-only modules, the form most consistent with deliberate source omission and 
the subset warranting closest inspection.}

%% file: sections/5-evaluation/2_RQ2.tex
\subsection{RQ2: Practical Analyzability}\label{sec:RQ2}

RQ2 evaluates whether the bytecode observed in PyPI distributions
is practically analyzable using a version-aware analysis stack.
The completed RQ2 run covers bytecode that resolves to CPython 3.8--3.14, 
yielding 204{,}904 in-scope \texttt{.pyc} files across 6{,}784 artifacts with CPython~3.8--3.14 bytecode.
RQ2 uses the same version criterion as Table~\ref{tab:bytecode-versions}. 
The RQ2 scope excludes bytecode assigned to older CPython versions, PyPy, unknown or unsupported versions, and the singleton CPython 3.15 file, because those files cannot be mapped to the prepared modern CPython analysis environments.

\subsubsection{Analysis Results}

Within the RQ2 scope, 183{,}363 files (89.49\%) have a corresponding source path in the same artifact, 
while 21{,}541 files (10.51\%) are artifact-local source-less \texttt{.pyc} files under the artifact-local source check. 
Table~\ref{tab:tool-analyzability} reports the strongest successful analysis level for each \texttt{.pyc} file, 
and Table~\ref{tab:rq2-tool-outcomes} reports the underlying tool outcomes.

\begin{table}[htp]
  \centering
  \caption{Distribution of bytecode analyzability levels by CPython version. The version counts use the same criterion as Table~\ref{tab:bytecode-versions}; percentages are calculated within each version.}
  \label{tab:tool-analyzability}
  \begin{tabular}{lrrrrrr}
    \toprule
    Version & \texttt{.pyc} files & L4 & L3 & L2 & L1 & L0 \\
    \midrule
    CPython 3.8 & 31{,}633 & 31{,}631 (99.994\%) & 2 (0.006\%) & 0 & 0 & 0 \\
    CPython 3.9 & 30{,}321 & 30{,}320 (99.997\%) & 1 (0.003\%) & 0 & 0 & 0 \\
    CPython 3.10 & 51{,}487 & 51{,}487 (100.000\%) & 0 & 0 & 0 & 0 \\
    CPython 3.11 & 38{,}320 & 38{,}320 (100.000\%) & 0 & 0 & 0 & 0 \\
    CPython 3.12 & 36{,}637 & 36{,}637 (100.000\%) & 0 & 0 & 0 & 0 \\
    CPython 3.13 & 13{,}274 & 13{,}274 (100.000\%) & 0 & 0 & 0 & 0 \\
    CPython 3.14 & 3{,}232 & 3{,}232 (100.000\%) & 0 & 0 & 0 & 0 \\
    \midrule
    Total & 204{,}904 & 204{,}901 (99.999\%) & 3 (0.001\%) & 0 & 0 & 0 \\
    \bottomrule
  \end{tabular}
\end{table}

Table~\ref{tab:tool-analyzability} shows that practical analyzability 
is effectively universal for the modern CPython bytecode observed in PyPI. 
PyLingual successfully decompiles 204{,}901 of 204{,}904 in-scope files, 
allowing nearly all files to reach L4, meaning that at least one selected decompiler emits source. 
{As defined in Section~\ref{sec:method:source}, L4 records successful source emission, not 
verified functional equivalence; the high L4 rate therefore establishes that modern PyPI bytecode is 
broadly amenable to source emission by current tools, not that recovered source faithfully reproduces 
the original behavior.}
Only three files fail to reach L4, 
and all three remain at L3 under the final level assignment. 
No file remains only loadable, only disassemblable, or completely unanalyzable under the final level assignment. 
The rare L3 cases occur in CPython 3.8 and 3.9 artifacts; all CPython 3.10--3.14 files reach L4.
A particularly important result concerns artifact-local source-less \texttt{.pyc} files. 
Among the 21{,}541 in-scope CPython~3.8--3.14 artifact-local source-less \texttt{.pyc} files, all reach L4, meaning that at least one selected decompiler emits source. 
These files appear in 1,387 artifacts,
including 562 artifacts exhibiting dynamic-loading indicators
and 64 artifacts matching security-relevant keyword categories.
Thus, 
artifact-local source-less \texttt{.pyc} files are not inherently opaque to modern analysis tooling, 
even though it remains invisible to source-only package inspection. 
This distinction is important because package-security workflows that rely exclusively on source files may overlook executable content 
that is nevertheless recoverable through bytecode-aware analysis.

Table~\ref{tab:rq2-tool-outcomes} provides additional insight 
into how analyzability is achieved. 
CPython reference loading succeeds for 204{,}551 files (99.828\%), 
and disassembly succeeds for 204{,}506 of the files that load (99.978\%). 
PyLingual succeeds on 204{,}901 files (99.999\%).
As an independent native decompiler check, Decompyle++ succeeds on 204{,}207 files (99.660\%) and returns controlled failure diagnostics for 697 files (0.340\%), with no timeouts.
However, 
the tool-level results indicate that failures are concentrated in individual analysis components 
rather than in the bytecode itself. 
Across all evaluated analysis tools, 1{,}098 tool failures are observed, including 697 controlled Decompyle++ rejections. 
Only three files fail to reach L4 in the final level assignment. 
The practical implication is that analyzability 
should be assessed using complementary tools rather than relying on a single loader, disassembler, or decompiler.

\begin{table}[htp]
  \centering
  \caption{Tool-level outcomes for bytecode analysis. \emph{Measured files} denotes the number of files processed by each tool; success, non-timeout failure, and timeout are additive outcome columns that partition those files. CPython \texttt{dis} is evaluated only after successful code-object loading via \texttt{marshal}.}
  \label{tab:rq2-tool-outcomes}
  \begin{tabular}{lrrrr}
    \toprule
    Tool & Measured files & Success & Failure & Timeout \\
    \midrule
    \texttt{marshal} & 204{,}904 & 204{,}551 (99.828\%) & 353 (0.172\%) & 0 \\
    \texttt{dis} & 204{,}551 & 204{,}506 (99.978\%) & 45 (0.022\%) & 0 \\
    \texttt{Decompyle++} & 204{,}904 & 204{,}207 (99.660\%) & 697 (0.340\%) & 0 \\
    PyLingual & 204{,}904 & 204{,}901 (99.999\%) & 0 & 3 (0.001\%) \\
    \bottomrule
  \end{tabular}
\end{table}

\subsubsection{Failed-Case Inspection}\label{ss:failinspection}

We record each failed tool outcome together with the package, artifact, \texttt{.pyc} path, 
bytecode tag, final analyzability level, tool status, and failure reason. 
Table~\ref{tab:rq2-failure-categories} summarizes the observed failures. 
Across all evaluated analysis tools, 1{,}098 failed tool results are recorded. 
Most failures are controlled rejection outcomes rather than robustness failures: 
697 are Decompyle++ rejections, 351 are CPython \texttt{marshal} rejections, and only the three PyLingual timeouts reduce the final analyzability level to L3.

\begin{table}[htp]
  \centering
  \caption{{Tool-level failed outcomes. Counts are failed tool-result counts, not unique \texttt{.pyc} files. Complete stdout/stderr traces are retained only for tool robustness failures.}}
  \label{tab:rq2-failure-categories}
  \begin{tabular}{lrrrr}
    \toprule
    Tool & Failed Items & Expected rejection & Translation failure & Robustness failure \\ 
    \midrule
    \texttt{marshal} & 353 & 351 & 0 & 2 \\ 
    \texttt{dis} & 45 & 0 & 0 & 45 \\ 
    \texttt{Decompyle++} & 697 & 697 & 0 & 0 \\ 
    PyLingual & 3 & 0 & 0 & 3 \\ 
    \midrule
    Total & 1{,}098 & 1{,}048 & 0 & 50 \\ 
    \bottomrule
  \end{tabular}
\end{table}

We classify each failed tool result into one of three exclusive outcome categories. 
\emph{Expected rejection} denotes fail-closed bytecode loading or decompiler rejection without an 
uncaught traceback. \emph{Robustness failure} denotes a timeout or an unhandled internal exception 
(\texttt{IndexError}, \texttt{SystemError}, \texttt{ValueError}, \texttt{EOFError}, or a trace-backed 
\texttt{AssertionError}). \emph{Translation failure} denotes source-like output that does not preserve 
the original bytecode behavior; it is necessarily zero in Table~\ref{tab:rq2-failure-categories} 
because RQ2 does not exercise the decompile--compile--rerun workflow, which is measured separately 
in RQ4.
The failure categories reflect different limitations. 
\texttt{marshal} failures indicate that the resolved CPython interpreter could not load the payload 
as a code object, potentially due to malformed content, embedded non-code data, or residual 
version mismatches. 
\texttt{dis} failures occur after successful loading and therefore represent a narrower boundary 
where a loaded code object cannot be rendered through standard disassembly. 
Decompyle++ failures are controlled diagnostics rather than timeouts, uncaught tracebacks, or signal-terminated subprocesses.
PyLingual failures are limited to three timeouts, which are the only failures that reduce the final analyzability level.
Failed items are attributed to the pipeline stage that reported the failed tool result. 
During replay with full traceback capture, 
a small number of standard-library \texttt{dis} items were found to fail during replayed marshal loading before reaching the disassembly routine; 
we retain their pipeline attribution while reporting the replayed root location in Table~\ref{tab:rq2-robustness-failures}.

\begin{table}[htp]
  \centering
  \caption{{Tool robustness failures by tool, version, and bug type. Counts are tool-result counts from observed PyPI bytecode; the unique column reports stack-aware failure signatures after replaying standard-library failures.}}
  \label{tab:rq2-robustness-failures}
  \begin{tabular}{@{}lllrr@{}}
    \toprule
    Tool & Version & Bug type & Count & Unique \\
    \midrule
    \multirow[t]{3}{*}{\texttt{dis}}
      & CPython 3.10 & \texttt{IndexError}: \texttt{dis.\_get\_const\_info} & 42 & 1 \\
      & CPython 3.12 & \texttt{ValueError}: replayed marshal-load failure & 1 & 1 \\
      & CPython 3.14 & \texttt{ValueError}: replayed marshal-load failure & 2 & 1 \\
    \midrule
    \texttt{marshal}
      & CPython 3.9 & \texttt{EOFError}: replayed marshal-load failure & 2 & 1 \\
    \midrule
    \multirow[t]{2}{*}{\texttt{PyLingual}}
      & CPython 3.8 & Timeout & 2 & 1 \\
      & CPython 3.9 & Timeout & 1 & 1 \\
    \midrule
    \texttt{Decompyle++}
      & CPython 3.8--3.14 & No robustness failure & 0 & 0 \\
    \midrule
    Total & & & 50 & 6 \\
    \bottomrule
  \end{tabular}
\end{table}

Table~\ref{tab:rq2-robustness-failures} isolates the robustness-failure class most relevant to tool reliability. 
The 50 robustness-relevant tool results collapse to 6 stack-aware failure signatures. 
They comprise 45 CPython \texttt{dis} failures, two \texttt{marshal} \texttt{EOFError} failures, 
and three PyLingual timeouts. 
Replaying the standard-library failures with 
full traceback capture shows that the 42 CPython 3.10 \texttt{dis} outcomes share the same failing 
location in \texttt{dis.\_get\_instructions\_bytes} and \texttt{dis.\_get\_const\_info}, whereas the 
CPython 3.12 and 3.14 \texttt{dis} cases fail during replayed marshal loading before reaching \texttt{dis}. 
Decompyle++ returns 697 controlled \texttt{exit\_1} diagnostics, with no timeout, traceback, or 
signal-terminated subprocess. 
These failures show that bytecode-analysis tools can hang or throw uncaught exceptions on real PyPI bytecode 
even when another tool emits source for the same file. 
Because these tools process untrusted bytecode without executing it, 
the failures are  
anti-analysis robustness failures rather than mere implementation defects; 
we discuss their security 
relevance in Section~\ref{sec:tooling-implications}.
The three PyLingual timeouts are the only failures that reduce the final analyzability level.

From a security-analysis perspective, the primary concern is the gap between source-level visibility 
and bytecode-level behavior. Among all failed tool outcomes, 261 occur on artifact-local source-less \texttt{.pyc} files, and the 
in-scope artifact-local source-less \texttt{.pyc} set contains artifacts exhibiting dynamic-loading indicators. A source-only audit would 
either overlook these files or treat them as opaque binary content. 
The lesson from the failed cases is not that modern PyPI bytecode is difficult to analyze, but that 
practical analyzability depends on version-aware execution, complementary tools, and explicit failure 
classification. A single-tool workflow would incorrectly classify some analyzable files as opaque, and 
{would also hide robustness failures in the analysis tools themselves.} 
Consequently, package-security workflows should retain bytecode-specific failure logging, copies of 
failed inputs, and tool-diversity checks even when overall analyzability rates appear near perfect.

\subsubsection{Failed-Case Examples}

Table~\ref{tab:rq2-failure-cases} presents representative failed cases. 
The final level is the highest level reached by any selected analysis path, 
not the level of the failed tool. 
These cases are not vulnerabilities; rather, 
they illustrate how different failure modes arise and why a file 
can remain highly analyzable even when one analysis tool fails.
These examples refine the aggregate results.
Although modern CPython bytecode is overwhelmingly analyzable with current tooling,
analyzability depends on version-aware execution,
tool diversity, and explicit failure accounting.
A single aggregate L4 percentage would conceal cases where a standard-library component,
an independent native decompiler,
or a modern decompiler fails on the same artifact while alternative analysis paths remain successful.

\begin{table}[t]
  \centering
  \caption{Representative failed cases illustrating distinct analysis-tool failure modes. Final level denotes the highest level reached after fallback tools are considered.}
  \label{tab:rq2-failure-cases}
  \begin{tabular}{p{0.1\linewidth}p{0.2\linewidth}p{0.16\linewidth}p{0.1\linewidth}p{0.26\linewidth}}
    \toprule
    Package & Failed input & Failed tool/result & Final level & What it shows \\ 
    \midrule
    \makecell[tl]{\texttt{dovado-rtl}\\ 0.10.12} 
    & generated SystemVerilog parser, CPython 3.9 
    & PyLingual timeout & L3 
    & Large generated parser bytecode can stress source recovery and leave the file below L4. \\ 
    
    \makecell[tl]{\texttt{sceleto2}\\ 1.1.2} 
    & \texttt{data.cpython-38.pyc} in both wheel and source distribution 
    & PyLingual timeout & L3 
    & The same timeout pattern appears in multiple distribution artifacts, so package-level grouping is needed to avoid over-counting. \\ 
    
    \makecell[tl]{\texttt{rtx-deep}\\ 1.3.9} 
    & 42 CPython 3.10 files in one wheel 
    & \texttt{dis} \texttt{IndexError} & L4 
    & PyLingual decompiles the files, so reference disassembly failure does not necessarily imply source recovery failure. \\ 
    \bottomrule
  \end{tabular}
\end{table}

\find{\textbf{Answer to RQ2.}
Modern PyPI bytecode is overwhelmingly amenable to source emission: nearly all
in-scope files reach L4, where at least one decompiler emits source, though L4
does not imply verified functional equivalence. Robustness remains a separate
limitation: across unmodified PyPI bytecode, we observe 50 robustness-failure
tool results spanning \texttt{marshal}, \texttt{dis}, and \texttt{PyLingual}
(six signatures, dominated by \texttt{dis}), caused by hangs or uncaught
exceptions; \texttt{Decompyle++} produces only controlled rejections. Reliable
analysis therefore requires tool diversity and explicit failure accounting.}

%% file: sections/5-evaluation/3_RQ3.tex
\subsection{RQ3: Security-Relevant CPython Runtime Robustness}\label{sec:RQ3}

This experiment evaluates CPython robustness when executing malformed bytecode. 
The claim is an existence claim: 
the experiment asks whether the bytecode execution boundary is crash-reachable across CPython 3.8--3.14, 
not whether a single campaign estimates the stable number of crashes or ranks versions by security. 
Once a crash input is preserved and reproduced under the matching CPython build, 
additional fuzzing runs may find more or fewer groups but do not invalidate the demonstrated reachability. 
We therefore interpret the results as lower-bound evidence of crash reachability under bounded campaigns.
The study covers the same modern CPython versions analyzed previously (3.8--3.14). 
For each version, 
the pipeline extracts source programs from the corresponding CPython unittest suite, 
compiles them into \texttt{.pyc} seeds using the matching interpreter, 
builds an instrumented CPython release, and executes a 24-hour honggfuzz campaign with a single worker. 
Each mutated \texttt{.pyc} file is executed only by the CPython version that produced the seed corpus.

\subsubsection{Finding Overview}

Table~\ref{tab:runtime-results-rq3} summarizes the completed campaigns. 
Across seven CPython versions, the pipeline generated 103{,}456 bytecode seeds and executed seven 24-hour fuzzing campaigns. 
Honggfuzz retained 12{,}404 crash-triggering inputs and no timeout-triggering inputs~\cite{klees2018evaluating}. 
Stack-based deduplication reduced these crashes to 1{,}009 unique crash groups. These counts are observed outcomes from one campaign per version and should be read as lower-bound reachability evidence rather than distributional estimates.

\begin{table}[htp]
  \centering
  \caption{CPython bytecode fuzzing results. Each campaign uses a single honggfuzz worker for 24 hours. Unique crash groups are deduplicated using reproduced native stacks when available and the honggfuzz stack hash otherwise.}
  \label{tab:runtime-results-rq3}
  \begin{tabular}{lrrrrr}
    \toprule
    Version & Seeds & Duration & Crash findings & Timeouts & Unique groups \\
    \midrule
    CPython 3.8 & 13{,}841 & 24h & 729 & 0 & 112 \\
    CPython 3.9 & 14{,}462 & 24h & 611 & 0 & 109 \\
    CPython 3.10 & 15{,}175 & 24h & 897 & 0 & 127 \\
    CPython 3.11 & 15{,}080 & 24h & 2{,}461 & 0 & 144 \\
    CPython 3.12 & 14{,}479 & 24h & 2{,}276 & 0 & 156 \\
    CPython 3.13 & 14{,}811 & 24h & 3{,}342 & 0 & 198 \\
    CPython 3.14 & 15{,}608 & 24h & 2{,}088 & 0 & 163 \\
    \midrule
    Total & 103{,}456 & 168h & 12{,}404 & 0 & 1{,}009 \\
    \bottomrule
  \end{tabular}
\end{table}

Malformed bytecode triggers abnormal native termination in every evaluated CPython version. 
Although the number of crashes varies across versions, 
these differences should not be interpreted as a direct ranking of security-relevant runtime robustness 
because fuzzing yield depends on seed composition, mutation trajectories, 
and version-specific bytecode semantics. 
The supported conclusion is narrower: 
across all tested modern releases, the bytecode execution boundary 
remains reachable by inputs that induce native-process failures.
The deduplication stage replays crash inputs under \texttt{gdb} 
using the corresponding instrumented CPython build and version-specific runtime environment~\cite{van2018semantic}. 
For each unique group, 
it records the signal, stack signature, representative input, 
and representative \texttt{.pyc} file. Of the 1{,}009 unique groups, 956 are confirmed through reproduced \texttt{gdb} stacks, 
while 53 are grouped using honggfuzz stack hashes when replay does not recover a native stack. 
After signal normalization, 917 groups terminate with \texttt{SIGSEGV} and 92 with \texttt{SIGABRT}.

\begin{table}[htp]
  \centering
  \caption{Failure types for stack-deduplicated crash groups, grouped by symptom severity. 
  Counts are unique crash groups; findings are raw crash inputs assigned to those groups. 
  The severity grouping is symptom-based and does not establish exploitability. 
  Heap allocator aborts are split by glibc abort message: of 58 groups, 56 carry an explicit 
  corruption or invalidity diagnostic (e.g., \texttt{corrupted size vs.\ prev\_size}, 
  \texttt{free(): invalid pointer}, \texttt{double free or corruption}), including 
  unaligned-chunk detections and glibc malloc assertions that likewise indicate already-corrupted 
  allocator state, and are placed in the memory-corruption class; the remaining 2 groups have no 
  captured abort message and are conservatively retained in the availability class. No allocator 
  abort corresponded to a clean out-of-memory condition.}
  \label{tab:rq3-bug-types}
  \begin{tabular}{lrr}
    \toprule
    Failure type & Unique groups & Findings \\
    \midrule
    \multicolumn{3}{l}{\textit{Class A: availability / anti-analysis}} \\
    \quad Null/near-null pointer dereference & 709 & 7{,}093 \\
    \quad Fatal-error abort & 32 & 2{,}899 \\
    \quad Heap allocator abort (no diagnostic) & 2 & 11 \\
    \quad Stack-recursion pattern & 3 & 11 \\
    \quad Other abort signal & 2 & 2 \\
    \quad\textit{Subtotal} & \textit{748} & \textit{10{,}016} \\
    \midrule
    \multicolumn{3}{l}{\textit{Class B: potential memory-corruption}} \\
    \quad Invalid pointer dereference & 195 & 1{,}744 \\
    \quad Heap allocator abort (corruption diagnostic) & 56 & 217 \\
    \quad Invalid unmapped access & 4 & 420 \\
    \quad Heap/lifetime corruption symptom & 6 & 7 \\
    \quad\textit{Subtotal} & \textit{261} & \textit{2{,}388} \\
    \midrule
    Total & 1{,}009 & 12{,}404 \\
    \bottomrule
  \end{tabular}
\end{table}

To characterize the observed failures, 
we classify each unique group using its signal, fault address, faulting instruction, 
and normalized stack prefix. 
Table~\ref{tab:rq3-bug-types} summarizes the resulting categories, 
grouped into two severity classes. 
These labels describe observed crash symptoms rather than root causes or exploitability. 
For example, 
near-null fault addresses suggest likely null-pointer dereference behavior, 
whereas allocator-related frames indicate potential heap-management failures. 
Determining exploitability requires manual debugging and memory-safety analysis.

To separate availability impact from potential memory-safety impact, 
we group the unique crash groups by symptom severity. 
We treat null/near-null dereferences, controlled aborts (fatal-error, allocator, 
and other abort signals), and stack-recursion patterns 
as an \emph{availability / anti-analysis} class (748 groups), 
since these manifest as crashes, aborts, or hangs that deny or evade inspection 
without evidencing memory corruption. 
We treat invalid (non-null) pointer dereferences, invalid unmapped accesses, 
and heap/lifetime corruption symptoms as an \emph{observed potential memory-corruption} class (261 groups in these campaigns), 
where the faulting address is attacker-influenced 
or the symptom indicates allocator-state corruption. 
This grouping is symptom-based and does not establish exploitability, 
which would require minimization and manual memory-safety analysis; 
we use it only to route the stronger interpretation toward the smaller corruption-class subset 
and to avoid attributing memory-safety significance to availability-class crashes. 
For the load-but-do-not-execute consumers identified in the threat model (Section~\ref{sec:threat}), 
both classes are relevant---the availability class as an anti-analysis primitive 
and the corruption class as a candidate for escalation---whereas 
under the ordinary import path both reduce to denial of service.

We caution against reading the per-version crash counts as a security ranking. 
Abort-based failures are less common by unique-group count 
but account for a substantial share of raw findings 
because several groups are rediscovered repeatedly during fuzzing. 
Raw crash counts therefore primarily reflect fuzzer productivity, 
whereas stack-deduplicated groups more closely approximate 
the number of distinct runtime robustness issues requiring further investigation. 
Collectively, the results show that malformed bytecode 
can exercise failure-prone execution paths throughout the modern CPython runtime, 
and that an observed minority but non-trivial subset (261 groups in this run) exhibits 
memory-corruption symptoms warranting prioritized triage.

\subsubsection{Per-Version Finding Distributions}

Table~\ref{tab:rq3-bug-types-version} breaks the observed crash symptoms down by CPython version. 
Null or near-null pointer dereferences are the dominant failure category across all versions, 
accounting for the majority of unique crash groups. 
However, 
the distribution is not uniform. 
CPython 3.11--3.14 exhibit substantially more invalid-pointer and allocator-related symptoms than CPython 3.8--3.10, 
while CPython 3.13 produces the largest number of unique crash groups. 
These differences suggest that mutated bytecode 
reaches different runtime states and failure paths across interpreter generations. 
They should not be interpreted as a direct ranking of security-relevant runtime robustness 
because fuzzing yield depends on mutation trajectories, bytecode semantics, and implementation changes between releases.

\begin{table}[htp]
  \centering
  \caption{Failure types by CPython version. \emph{Allocator/heap} combines heap allocator aborts and heap/lifetime corruption symptoms; \emph{Other} combines invalid unmapped access, stack-recursion patterns, and other abort signals. The per-version columns use a coarser grouping than the severity split in Table~\ref{tab:rq3-bug-types}: \emph{Allocator/heap} spans both severity classes (allocator aborts are Class~A; heap/lifetime corruption is Class~B), so this table is not a per-version decomposition of the Class~A/Class~B totals.}
  \label{tab:rq3-bug-types-version}
  \begin{tabular}{lrrrrrr}
    \toprule
    Version & Unique & Near-null & Invalid ptr. & Fatal abort & Allocator/heap & Other \\
    \midrule
    CPython 3.8 & 112 & 100 & 4 & 6 & 1 & 1 \\
    CPython 3.9 & 109 & 88 & 10 & 2 & 8 & 1 \\
    CPython 3.10 & 127 & 112 & 8 & 1 & 5 & 1 \\
    CPython 3.11 & 144 & 73 & 53 & 7 & 9 & 2 \\
    CPython 3.12 & 156 & 89 & 47 & 7 & 13 & 0 \\
    CPython 3.13 & 198 & 130 & 42 & 4 & 20 & 2 \\
    CPython 3.14 & 163 & 117 & 31 & 5 & 8 & 2 \\
    \bottomrule
  \end{tabular}
\end{table}

To better understand where failures occur, 
Table~\ref{tab:rq3-context-version} classifies unique crash groups 
according to the runtime context inferred from their stack prefixes. 
We use this classification to separate failures localized to the bytecode-ingestion path---code-object 
loading and marshal deserialization, which is the surface CPython documents as unsafe for untrusted 
input~\cite{python-marshal}---from failures that occur only after mutated bytecode has survived 
ingestion and reached interpreter execution. 
Object-runtime operations constitute the largest category overall, 
followed by frame evaluation and garbage-collection or finalization activities. 
Code-object loading and serialization account for only 25 of the 1{,}009 groups (2.5\%), 
while a further 59 groups (5.8\%) cannot be attributed to a specific context and are recorded as unknown. 
In these campaigns, the remaining 925 groups (91.7\% of observed unique groups) crash in post-ingestion execution contexts.

\begin{table}[t]
  \centering
  \caption{Runtime contexts by CPython version. \emph{Code object} combines code-object loading and serialization contexts. For the loading-vs-execution split discussed in the text, \emph{Code object} and \emph{Unknown} are treated as ingestion-path or unattributable (84 groups total), while \emph{Object runtime}, \emph{Frame eval.}, \emph{GC/final.}, and \emph{Instrument.} are post-ingestion execution contexts (925 groups). Per-version differences reflect the observed mutation trajectories, seed composition, and bytecode semantics of this run, not relative interpreter security.}
  \label{tab:rq3-context-version}
  \begin{tabular}{lrrrrrrr}
    \toprule
    Version & Unique & Object runtime & Frame eval. & GC/final. & Instrument. & Code object & Unknown \\
    \midrule
    CPython 3.8 & 112 & 46 & 48 & 11 & 0 & 1 & 6 \\
    CPython 3.9 & 109 & 42 & 37 & 15 & 0 & 0 & 15 \\
    CPython 3.10 & 127 & 43 & 60 & 16 & 0 & 0 & 8 \\
    CPython 3.11 & 144 & 93 & 21 & 25 & 0 & 0 & 5 \\
    CPython 3.12 & 156 & 92 & 12 & 31 & 9 & 7 & 5 \\
    CPython 3.13 & 198 & 120 & 21 & 27 & 8 & 12 & 10 \\
    CPython 3.14 & 163 & 109 & 25 & 10 & 4 & 5 & 10 \\
    \bottomrule
  \end{tabular}
\end{table}

The context distribution indicates 
that malformed bytecode is not confined to parser- or loader-level failures. 
Some crash groups originate during code-object construction and deserialization, 
with representative stacks passing through routines 
such as \texttt{marshal\_loads}, \texttt{r\_object}, \texttt{\_PyCode\_New}, and quickening-related helpers. 
Others occur after execution begins, 
involving frame evaluation, object manipulation, attribute lookup, 
iterator processing, instrumentation support, garbage collection, 
and interpreter finalization. 
The quantitative split is the key observation: at most 84 groups (8.3\%) fall in the 
loading or unknown contexts, so at least 925 groups (91.7\%) reach interpreter execution 
beyond the ingestion boundary. 
Because the 59 unknown groups cannot be assigned to either side, we report the 
execution-phase share as a lower bound rather than an exact split.

Taken together, 
these results show that bytecode-induced failures are distributed 
across multiple phases of CPython execution 
rather than being concentrated at the bytecode-loading boundary. 
The observed crashes span code-object construction, execution, 
runtime object management, instrumentation, serialization, and cleanup, 
suggesting that malformed bytecode can propagate 
through substantial portions of the interpreter before triggering abnormal termination. 
This distinction matters for interpretation. 
Crashes localized to marshal deserialization largely confirm a documented non-guarantee, 
since marshal is explicitly not secure against malformed or malicious data~\cite{python-marshal}; 
the more notable result is empirical: the overwhelming majority of distinct findings 
arise only after mutated bytecode survives initial ingestion and reaches core interpreter execution. 
Modern CPython runtimes therefore remain crash-reachable through malformed bytecode inputs, 
and the resulting failures extend well beyond bytecode ingestion 
into core interpreter execution and runtime-management logic.

\subsubsection{Case Studies}

Table~\ref{tab:rq3-case-studies} presents representative stack-deduplicated crash groups. 
The examples are selected to illustrate distinct interpreter subsystems reached by malformed bytecode 
rather than to rank findings by severity or frequency.

\begin{table}[htp]
  \centering
  \caption{Representative crash groups. The stack prefix lists the first distinctive function frames in the reproduced stack signature.}
  \label{tab:rq3-case-studies}
  \begin{tabular}{p{0.14\linewidth}p{0.08\linewidth}p{0.28\linewidth}p{0.34\linewidth}}
    \toprule
    Version & Findings & Stack prefix & Interpretation \\
    \midrule
    CPython 3.13  & 861 & \texttt{\_PyEval\_EvalFrameDefault} & A high-frequency segmentation-fault group reached frame evaluation before crashing; the unresolved top frame requires manual triage. \\
    CPython 3.13  & 91 & \texttt{deopt\_code}, \texttt{\_PyCode\_GetCode}, \texttt{marshal\_dumps} & Mutated bytecode can reach code-object serialization and deoptimization paths, not only direct execution. \\
    CPython 3.13  & 63 & \texttt{\_Py\_GetBaseOpcode}, \texttt{\_PyCode\_Quicken}, \texttt{\_PyCode\_New} & Some crashes occur while loading mutated bytecode into code objects and applying quickening-related logic. \\
    CPython 3.13  & 8 & \texttt{get\_tools\_for\_instruction}, instrumentation callbacks & Mutated instruction streams can reach newer interpreter instrumentation paths. \\
    \bottomrule
  \end{tabular}
\end{table}

These examples highlight two important observations. 
First, 
stack-based deduplication is essential for interpreting fuzzing results. 
The 12{,}404 retained crash inputs collapse to 1{,}009 unique crash groups, 
and several groups are rediscovered hundreds of times. 
Raw crash counts therefore reflect fuzzer productivity, 
whereas stack-deduplicated groups better approximate the number of distinct runtime failures requiring investigation.
Second, 
the affected execution contexts are diverse. 
The representative groups span frame evaluation, code-object construction, quickening, serialization, deoptimization, 
and instrumentation support. 
Together with the distributions reported earlier, 
these cases indicate that malformed bytecode is not confined to a single validation boundary or opcode handler. 
Instead, mutated bytecode can propagate through multiple stages of interpreter processing before triggering abnormal termination.

From a security perspective, 
the significance of these findings lies in reachability rather than exploitability. 
The observed crashes demonstrate that malformed bytecode 
can drive execution into a broad set of internal CPython subsystems. 
However, these findings should be interpreted as runtime robustness failures 
rather than confirmed vulnerabilities. 
Establishing exploitability would require additional minimization, 
root-cause analysis, memory-safety debugging, and evaluation against CPython's intended trust assumptions for bytecode execution.

\find{\textbf{Answer to RQ3.} 
Mutated bytecode can trigger abnormal native termination across all evaluated CPython versions (3.8--3.14), producing a diverse set of stack-deduplicated runtime failures dominated by pointer-dereference symptoms, an observed smaller subset of which (261 groups in this campaign set) exhibits potential memory-corruption characteristics. 
These findings show that malformed bytecode can reach deep interpreter
execution paths and that bytecode execution remains a robustness boundary
requiring systematic analysis and triage.
}

%% file: sections/5-evaluation/4_RQ4.tex
\subsection{RQ4: Source Reproduction}\label{sec:RQ4}

RQ4 evaluates whether the bytecode-level runtime findings identified in RQ3 
can be translated back into ordinary Python source while preserving the observed behavior. 
Specifically, 
we examine whether source recovered by version-matched decompilers 
can be recompiled and used to reproduce the original bytecode-level finding. 
Each of the 1{,}009 unique crash groups is analyzed 
using PyLingual and Decompyle++ under the matching CPython version. 
A finding is counted as source-reproduced only if recovered source compiles 
under the same interpreter and reaches the same behavior class as the original mutated input.

\subsubsection{Source-Reproduction Results}
Table~\ref{tab:source-reproduction} summarizes the finding-level outcomes using exclusive source-reproduction stages. 
None of the 1{,}009 stack-deduplicated runtime findings can be reproduced through recovered Python 
source. Although the bytecode findings are analyzable and often partially decompilable, no recovered 
source program recompiles into bytecode that reproduces the original runtime behavior. 
The failures are concentrated before a usable source reproducer exists: 662 findings produce no source reproducer, 
285 produce source-like output that fails to compile, and 62 compile but do not reproduce the original 
bytecode behavior on source rerun. These finding-level categories map directly onto the tool-level results in Table~\ref{tab:rq4-tool-failures}: the 285 compile failures and 62 non-reproducing reruns are the 347 Decompyle++ source compile/rerun failures reported there, while the four Decompyle++ robustness failures remain in the no-source-reproducer group. Because the inputs are the stack-deduplicated 
findings already confirmed in RQ3, we classify outcomes by source-recovery stage rather than 
introducing a separate replay-confirmation category.

\begin{table}[htp]
  \centering
  \caption{Tool-bounded source reproduction for RQ3 findings. Categories are finding-level and exclusive. Source reproduction requires recovered source to compile under the matching CPython version and reach the same behavior class as the original bytecode.}
  \label{tab:source-reproduction}
  \begin{tabular}{@{}lrrrrr@{}}
    \toprule
    Version & Findings & \makecell{No source\\reproducer} & \makecell{Source\\compile fail} & \makecell{Source rerun\\not reproduced} & \makecell{Source\\reproduced} \\
    \midrule
    CPython 3.8 & 112 & 75 & 18 & 19 & 0 \\
    CPython 3.9 & 109 & 69 & 19 & 21 & 0 \\
    CPython 3.10 & 127 & 75 & 50 & 2 & 0 \\
    CPython 3.11 & 144 & 86 & 50 & 8 & 0 \\
    CPython 3.12 & 156 & 67 & 82 & 7 & 0 \\
    CPython 3.13 & 198 & 127 & 66 & 5 & 0 \\
    CPython 3.14 & 163 & 163 & 0 & 0 & 0 \\
    \midrule
    Total & 1{,}009 & 662 & 285 & 62 & 0 \\
    \bottomrule
  \end{tabular}
\end{table}

The distribution remains asymmetric across versions. CPython 3.14 findings are blocked before source compilation, 
whereas CPython 3.8--3.13 include cases where source-like output is emitted but fails to compile or compiles without reproducing 
the original bytecode behavior. This does not mean those findings are source-reproducible; it reflects cases where 
translation advances far enough to expose compilation or behavioral-preservation failures. 
More broadly, the results highlight a fundamental distinction between bytecode analyzability and 
source-level reproducibility. A bytecode artifact may be successfully inspected, disassembled, or 
partially decompiled while still failing to yield source that preserves the behavior responsible for 
the original runtime finding.

\subsubsection{Why Source Reproduction Fails}\label{sec:rq4-why}

Table~\ref{tab:rq4-tool-failures} reports tool-level outcomes using the same three category names as Table~\ref{tab:rq2-failure-categories}: \emph{expected rejection}, \emph{translation failure}, 
and \emph{tool robustness failure}. In the RQ4 setting these denote, respectively, a fail-closed 
response before usable source is emitted, emitted source-like output that either fails to compile or compiles without reproducing the original bytecode 
behavior, and an uncaught exception, hang, timeout, or process-level crash or abort in the evaluated 
tool. Unlike the RQ2 failures, which were managed-code exceptions, the RQ4 robustness failures include 
native process terminations, as detailed below. 
The analysis generates 2{,}018 tool rows, one per combination of RQ3 finding and source-recovery tool. 
PyLingual fails closed for most findings, with 19 robustness failures. Decompyle++ rejects 658 findings before usable source is emitted, 
emits source-like output for 347 findings that either fails to compile or compiles without reproducing the original bytecode-level behavior, and exhibits four robustness failures. 
Across both tools, the robustness failures include uncaught exceptions, timeouts, and signal-terminated subprocesses.

\begin{table}[htp]
  \centering
  \caption{Tool-level outcomes. Counts are tool-result counts, not unique findings. Complete stdout/stderr traces are retained only for tool robustness failures.}
  \label{tab:rq4-tool-failures}
  \begin{tabular}{lrrrr}
    \toprule
    Tool & Failed rows & Expected rejection & Translation failure & Robustness failure \\
    \midrule
    PyLingual & 1{,}009 & 990 & 0 & 19 \\
    \texttt{Decompyle++} & 1{,}009 & 658 & 347 & 4 \\
    \midrule
    Total & 2{,}018 & 1{,}648 & 347 & 23 \\
    \bottomrule
  \end{tabular}
\end{table}

\begin{table}[t]
  \centering
  \caption{Decompiler robustness failures. Counts are tool-result counts; uncaught exceptions, timeouts, and signal-terminated subprocesses are all classified as robustness failures. The unique column reports stack-aware failure signatures when available and falls back to version-and-signal grouping when no native stack is retained.}
  \label{tab:rq4-trace-errors}
  \begin{tabular}{@{}lllrr@{}}
    \toprule
    Tool & Version & Failure type and cause & Count & Unique \\
    \midrule

    \multirow[t]{8}{*}{\texttt{PyLingual}}
      & \multirow[t]{2}{*}{CPython 3.8}
      & Timeout: did not terminate within 600s & 3 & 1 \\
      & & \texttt{SIGILL}: process crashed with signal 4 & 1 & 1 \\
    \cmidrule(lr){2-5}
      & \multirow[t]{2}{*}{CPython 3.9}
      & \texttt{SIGILL}: process crashed with signal 4 & 2 & 1 \\
      & & Timeout: did not terminate within 600s & 2 & 1 \\
    \cmidrule(lr){2-5}
      & \multirow[t]{2}{*}{CPython 3.10}
      & \texttt{SIGILL}: process crashed with signal 4 & 3 & 1 \\
      & & Timeout: did not terminate within 600s & 1 & 1 \\
    \cmidrule(lr){2-5}
      & \multirow[t]{2}{*}{CPython 3.12}
      & \texttt{SIGSEGV}: process crashed with signal 11 & 4 & 1 \\
      & & \texttt{SIGABRT}: allocator/heap abort & 3 & 1 \\

    \midrule

    \multirow[t]{3}{*}{\texttt{Decompyle++}}
      & CPython 3.8 & \texttt{SIGSEGV}: process crashed with signal 11 & 2 & 1 \\
      & CPython 3.11 & Timeout: did not terminate within 600s & 1 & 1 \\
      & CPython 3.13 & \texttt{SIGKILL}: process terminated with signal 9 & 1 & 1 \\

    \midrule
    Total & & & 23 & 11 \\
    \bottomrule
  \end{tabular}
\end{table}

Table~\ref{tab:rq4-trace-errors} summarizes the retained robustness failures
across the evaluated decompilers. The 23 tool-result failures collapse to 11
signatures after grouping by tool, CPython version, failure mode, and retained
diagnostic evidence; signal-terminated Decompyle++ rows with no native stack are
grouped conservatively by version and signal. These are tool robustness
limitations rather than ordinary source-translation failures: instead of
rejecting malformed bytecode through controlled diagnostics, the decompilers can
raise uncaught exceptions, hang, or terminate by native signal on a subset of
adversarial inputs.
Beyond bounding source reproduction, these failures are a concrete instance of the
load-but-do-not-execute threat (Section~\ref{sec:threat}). Although the decompilers
mostly reject the mutated inputs or emit source-like output that does not reproduce
the finding, 23 rows drive a tool into a data-channel failure when asked to inspect
adversarial bytecode rather than run trusted application code, showing that the
non-executing consumer class is not hypothetical. We do not claim these crashes are
exploitable, which would require per-tool minimization and memory-safety analysis.
Their significance is that an untrusted-bytecode consumer can be driven into native
process failure through a pure data channel, an availability and anti-analysis risk
independent of any code-execution capability the attacker may hold.

In summary, the tool-level results explain the absence of source-level reproduction. 
Source 
recovery is not binary: it depends on version support, decompiler translation depth, and whether the 
mutated bytecode remains representable in the source language the decompiler reconstructs. 
Here, deeper 
analysis does not yield reproduction. Decompyle++ emits source-like output for 347 findings; 285 fail to compile and 62 compile without reproducing the original bytecode-level finding. More importantly, the results reveal a clear distinction 
between bytecode-level and source-level reachability: although the RQ3 findings are observable and 
analyzable at the bytecode level, none reproduce through recovered Python source under the evaluated 
workflows. Bytecode-level behaviors can therefore persist beyond the representational limits of current 
source-recovery tools and should not automatically be read as source-level reachable.

\find{\textbf{Answer to RQ4.} 
None of the 1{,}009 confirmed bytecode-level runtime findings can be reproduced through recovered 
Python source using the evaluated source-recovery workflows, showing that bytecode analyzability does 
not imply source-level reproducibility and that bytecode-level behaviors can remain unreachable from 
ordinary source even when partial decompilation succeeds. 
The reproduction attempts also expose a robustness cost: 23 tool runs (11 distinct signatures) drive decompilers into uncaught exceptions, timeouts, or signal-terminated subprocess failures, confirming that untrusted bytecode can disrupt analysis tools through a pure data channel.}

%% file: sections/6-discussion/discussion.tex
\section{Discussion} \label{sec:diss}

The results reveal a consistent gap between package-level visibility,
bytecode-level behavior, and source-level interpretation. RQ1 shows that bytecode
is present in real package distributions, RQ2 shows that it is generally
analyzable with version-aware tooling, RQ3 shows that malformed bytecode can
trigger runtime failures across modern CPython releases, and RQ4 shows that those
failures are not reproduced through the evaluated source-recovery workflows. This
section discusses the implications of that gap for package security, runtime
hardening, analysis workflows, and responsible interpretation of bytecode-level
findings.

\subsection{Implications for Package Security}

The RQ1 results show why bytecode cannot be treated as an exceptional corner case
in package analysis. The PyPI scan finds 7{,}388 bytecode-containing distribution
artifacts and 228{,}578 \texttt{.pyc} files, including 28{,}193 artifact-local
source-less \texttt{.pyc} files; bytecode appears in both wheels and source
distributions, and 242{,}586 collected artifacts contain dynamic-loading
indicators. RQ2 refines this picture: artifact-local source-less \texttt{.pyc}
files are not opaque (all 21{,}541 in-scope CPython~3.8--3.14 such files reach L4
under PyLingual, i.e., source is emitted rather than semantically validated), yet
source-only inspection remains insufficient because it never learns that these
files require recovery in the first place. These results support three concrete
package-security requirements:

\begin{itemize}
  \item \textbf{Inventory bytecode explicitly.} Scanners should enumerate
  \texttt{.pyc} files, record their bytecode versions, preserve surrounding loader
  context, and distinguish artifact-local source-present from source-less files,
  since a source-only scan can miss executable content present in the distributed
  artifact.
  \item \textbf{Treat decompilation as recoverable, not sufficient.} Bytecode-aware
  analysis can often recover source-level views that source-only analysis would
  overlook entirely, so recovery should be a routine triage step rather than a
  fallback.
  \item \textbf{Retain per-tool failure evidence.} The 1{,}098 failed tool results
  in RQ2 show that pipelines should not collapse analysis into a single success
  flag: \texttt{marshal}, \texttt{dis}, Decompyle++, and PyLingual fail on
  different subsets, most affected files remain analyzable through another tool,
  and tool diversity with explicit failure accounting is therefore part of the
  security evidence.
\end{itemize}

\subsection{Implications for Python Runtime Security}

The strongest runtime-robustness implication of this study is that Python bytecode
should be treated as an explicit interpreter execution boundary. Although CPython
normally executes bytecode produced by its own compiler, the runtime also exposes
mechanisms that load serialized code objects and bytecode artifacts. RQ3 evaluates
this boundary directly: across CPython 3.8--3.14, seven 24-hour fuzzing campaigns
produce 12{,}404 retained crash inputs and 1{,}009 stack-deduplicated crash groups,
dominated by pointer-dereference failures and spanning frame evaluation,
object-runtime behavior, garbage collection, instrumentation, and code-object
loading or serialization. These findings carry three implications:

\begin{itemize}
  \item \textbf{Deserialization crashes confirm a documented non-guarantee.} The
  marshal loader is explicitly documented as not secure against erroneous or
  malicious data~\cite{python-marshal}, so the 25 of 1{,}009 groups (2.5\%)
  localized to code-object loading or serialization restate a known limitation
  rather than a new one.
  \item \textbf{Hardening should extend beyond the marshal boundary.} At least 925
  groups (91.7\%) reach interpreter execution past initial ingestion, spanning
  frame evaluation, object management, instrumentation, and finalization. Because
  failures are not confined to a single parser or opcode handler, validation should
  cover code-object and bytecode invariants checked before and during execution,
  particularly instruction streams, constant pools, exception tables, quickening
  metadata, and frame-state assumptions. We do not prove any specific invariant is
  unchecked, but the measured diversity of crash contexts makes the direction
  actionable.
  \item \textbf{Bytecode findings are not source-level vulnerabilities.} RQ4
  bounds the interpretation: none of the 1{,}009 findings reproduce from ordinary
  Python source (662 produce no reproducer, 285 produce source that fails to
  compile, 62 compile without reproducing the behavior), so a crash-triggering
  input identifies a hardening opportunity without showing that ordinary source can
  reach the same state.
\end{itemize}

\noindent For CPython implementers, the question is therefore not whether malformed
bytecode should be trusted, which it should not, but how the runtime should fail
when it is encountered through existing loading paths. A security-oriented runtime
should reject such inputs through predictable exceptions or structured rejection
mechanisms rather than native crashes, allocator aborts, or inconsistent
interpreter state. Robust rejection improves both interpreter hardening and the
safety of downstream package-security and malware-analysis workflows on untrusted
bytecode.

\subsection{Implications for Bytecode Analysis and Security Tooling}
\label{sec:tooling-implications}

The evaluation highlights complementary requirements for bytecode-aware security
analysis, separated by the kind of bytecode the tools face:

\begin{itemize}
  \item \textbf{On ecosystem bytecode, record level and failures together.} RQ2
  shows practical analyzability is high when analysis is version-aware: 204{,}901
  of 204{,}904 in-scope files reach L4 and only three remain at L3. But individual
  tools reject, time out, or disagree on the same artifact, and some hang or throw
  uncaught exceptions even on observed PyPI bytecode, so a robust workflow must
  record both the strongest successful level and the per-tool failure evidence,
  since a single failing tool would otherwise misclassify an analyzable file as
  opaque.
  \item \textbf{On adversarial bytecode, fail closed and explain why.} Both
  decompilers mostly reject mutated inputs, but neither fails closed completely:
  PyLingual produces 19 robustness failures and Decompyle++ four, collapsing to 11
  stack-aware signatures, and a subset terminate the process by native signal
  (\texttt{SIGSEGV}, \texttt{SIGABRT}, or \texttt{SIGILL}) rather than through a
  structured diagnostic. Tools should therefore provide explicit
  malformed-bytecode rejection paths, bounded execution, structured failure
  reasons, and clear separation between recovered source and diagnostic output.
\end{itemize}

\noindent These tool failures are themselves security-relevant for the
non-executing bytecode-analysis consumer. On observed PyPI bytecode (RQ2) the
failures are managed-code exceptions and timeouts rather than native-process
failures in our observations; on adversarial bytecode (RQ4) the same tools reach
native process failures, including fault classes observed in the
interpreter-execution experiment under RQ3. Across both settings the evaluated
tools exhibit 17 distinct robustness signatures (six on observed PyPI bytecode in
RQ2 and 11 under adversarial bytecode in RQ4), spanning PyLingual, Decompyle++, and
the CPython \texttt{dis} module. Because decompilers and disassemblers are run by
analysts and package-scanning pipelines on \texttt{.pyc} files of unknown
provenance, this matches the load-but-do-not-execute consumer of
Section~\ref{sec:threat}, in which the attacker controls a data channel rather than
a code channel:

\begin{itemize}
  \item \textbf{Availability primitive.} A timeout that reliably stalls a
  decompiler for 600 seconds is an availability primitive against a scanning
  pipeline.
  \item \textbf{Anti-analysis primitive.} An uncaught exception or crash that
  aborts analysis of a file yields no source-level verdict, which a malicious
  package could exploit to defeat source-recovery triage while remaining
  executable.
  \item \textbf{Already present, and escalating.} The risk is neither hypothetical
  nor confined to adversarial inputs: it manifests on bytecode present in PyPI
  today and escalates to native process failure under adversarial bytecode. The
  CPython \texttt{dis} \texttt{IndexError} observed in RQ2 is additionally a
  standard-library robustness failure on crafted constant-pool input and a
  candidate for upstream reporting under Section~\ref{sec:ethics}.
\end{itemize}

\noindent We do not claim these failures are exploitable for code execution, which
would require minimization and memory-safety analysis of each tool. The results
also bear on malware analysis: RQ1 shows bytecode and dynamic-loading indicators
appear in real package artifacts and RQ2 shows source-less \texttt{.pyc} files are
often recoverable with version-aware tooling, but decompilation alone is not
sufficient. A \texttt{.pyc} file is an executable interpreter input, not merely an
intermediate representation for source recovery, so bytecode-aware analysis should
combine version-aware loading, disassembly, code-object inspection, decompilation,
loader context, and tool-failure evidence as complementary views rather than
relying exclusively on recovered source.

\subsection{Ethical and Safety Considerations}
\label{sec:ethics}

This study analyzes public package artifacts and generated malformed bytecode
inputs, so the reporting boundary must be explicit. During data collection and
scanning, we do not install collected packages or import package modules: the
analysis operates on archive contents, file metadata, bytecode headers, extracted
code objects, generated seed corpora, and CPython builds, and external tools and
interpreter runs execute as bounded subprocesses. The study surfaces robustness
failures in two distinct classes of software, each warranting its own disclosure
path because they have separate maintainers and separate trust assumptions:

\begin{itemize}
  \item \textbf{The CPython interpreter (RQ3).} RQ3 reports crashes, aborts, and
  crash symptoms, but these are not automatically exploitable vulnerabilities, and
  RQ4 shows that the selected source-recovery workflows reproduce none of them from
  ordinary source. We therefore treat RQ3 as bytecode-level runtime evidence
  requiring triage. If a finding plausibly affects supported CPython releases or
  exposes a security-relevant failure mode after minimization and manual analysis,
  it should be reported through the upstream CPython disclosure process before any
  public release of triggering inputs or detailed reproduction artifacts.
  \item \textbf{Third-party analysis tools (RQ2, RQ4).} The tool findings are
  directly entailed by the load-but-do-not-execute threat model
  (Section~\ref{sec:threat}): an analyst running a decompiler or disassembler on
  untrusted bytecode controls a data channel into that tool. The concrete reporting
  candidates are the native-signal terminations observed in the decompilers under
  adversarial bytecode (the RQ4 signatures in Table~\ref{tab:rq4-trace-errors}, with
  \texttt{SIGSEGV}, \texttt{SIGABRT}, \texttt{SIGILL}, or \texttt{SIGKILL}) and the
  standard-library \texttt{dis} \texttt{IndexError} on crafted constant-pool input
  in RQ2 (Table~\ref{tab:rq2-failure-categories}). Each should be minimized to a triggering
  input and reported to the corresponding maintainer (the relevant decompiler
  repositories and, for \texttt{dis}, the CPython standard library), pinned to the
  tool version and commit on which we observed the failure so that triage is
  unambiguous and a later-version fix does not obscure the finding. Because
  Decompyle++ is evaluated from a fork rather than canonical pycdc, we report
  against the evaluated fork and check whether each failure reproduces upstream,
  notifying upstream maintainers where it does.
\end{itemize}

\noindent We characterize all of these as robustness and availability defects
rather than claimed-exploitable vulnerabilities; establishing exploitability would
require per-tool minimization and memory-safety analysis we do not perform. We
further distinguish the urgency of the tool findings by failure class:

\begin{itemize}
  \item \textbf{Memory-safe failures (RQ2).} Failures on bytecode already present
  in PyPI are managed-code exceptions and timeouts: anti-analysis but memory-safe,
  and reportable as ordinary robustness bugs without input-withholding.
  \item \textbf{Native-signal failures (RQ4).} Terminations observed under
  adversarial bytecode are handled with disclosure-first care, since a working
  crasher against a currently distributed analysis tool is the more sensitive
  artifact.
\end{itemize}

\noindent Finally, the artifact-release strategy reconciles reproducibility with
this disclosure discipline. We release {\tech}, the dataset-construction and
analysis scripts, and the manifests needed to reproduce the study, but we gate
rather than withhold the sensitive subset:

\begin{itemize}
  \item \textbf{Gated triggering inputs.} The specific crash-triggering bytecode
  inputs, both the CPython seeds-of-record for RQ3 and the adversarial inputs that
  terminate the analysis tools in RQ4, are held back from the public artifact until
  the corresponding disclosure windows have closed.
  \item \textbf{Dual-use mitigation.} The source-less \texttt{.pyc} characterization
  and the tool-crashing inputs could in principle aid an attacker targeting analysis
  pipelines; gated release of triggering inputs together with disclosure-first
  reporting is our mitigation, preserving auditability of the methodology and
  aggregate findings while avoiding distribution of working crashers against live
  interpreters and tools.
\end{itemize}

%% file: sections/7-threats/threats.tex
\section{Threats to Validity} \label{sec:threats}

This section summarizes the main threats to validity, their mitigations, and the
remaining scope after mitigation. The goal is not to eliminate every limitation,
but to keep claims tied to the populations, tools, and measurements that support
them.

\vspace{3pt}
\noindent
\textit{T1. Dataset Scope and Representativeness.}
The primary external-validity threat is dataset scope: the study measures PyPI
distribution artifacts rather than all Python deployments, so installed
environments, containers, application bundles, private indexes, and source
repositories are out of scope and may exhibit different bytecode exposure. We
mitigate this by selecting PyPI because it is the public package ecosystem consumed
by installers and package-security workflows and is widely used in prior empirical
security studies, and by fixing the denominator through a package-balanced policy
of at most one wheel and one source distribution per package. Consequently, the
results support claims about public PyPI distribution artifacts and should not be
generalized to all Python deployments.

\vspace{3pt}
\noindent
\textit{T2. Collection and Artifact Selection.}
The latest-release policy, five-year upload window, and one-wheel/one-source-distribution
rule shape which artifacts enter the dataset. These rules may omit historical
releases, older bytecode, and platform-specific wheel differences, and they
introduce survivorship bias: removed packages and deleted releases, including
removed malicious packages, are structurally absent. We mitigate this through
deterministic selection rules, reproducible manifests, and explicit artifact-type
tracking; the latest-release policy targets the versions most likely to be
retrieved and analyzed at collection time, and the five-year window aligns the
dataset with contemporary packaging and modern CPython/tool support. The resulting
dataset is therefore a contemporary package-balanced snapshot, not a longitudinal
history or a complete census of uploads, and RQ1 should not be read as a
measurement of where attackers hide bytecode.

\vspace{3pt}
\noindent
\textit{T3. Source Visibility and Bytecode Classification.}
Source visibility and bytecode classification introduce construct-validity threats,
since our checks are conservative and artifact-local. The artifact-local
source-less \texttt{.pyc} check establishes whether a plausible source counterpart
exists within the same artifact, not source-bytecode equivalence, and
dynamic-loading indicators are permissive textual matches over source-like archive
entries (for \texttt{marshal.}, \texttt{marshal.loads(}, \texttt{importlib},
\texttt{SourcelessFileLoader}, \texttt{exec(}, \texttt{eval(}, and
\texttt{types.CodeType}/\texttt{CodeType(}) that we did not validate through a
hand-labeled precision study. We mitigate these threats by reporting source
visibility, artifact-local source-less \texttt{.pyc} files, version evidence,
packaging structure, and dynamic-loading indicators as separate measurements rather
than collapsing them into a single transparency metric, since source presence,
decompilation success, and behavioral reproduction address different security
questions. Source visibility should therefore be read as transparency evidence
rather than proof of semantic correspondence, and dynamic-loading counts as an
upper-bound context signal rather than proof of runtime loading; the source-less
characterization (Section~\ref{sec:sourceless-char}) rests on packaging structure
and static topical keywords, so we report exposure and topical coincidence rather
than malicious behavior, leaving behavioral triage of the source-less set to future
work.

\vspace{3pt}
\noindent
\textit{T4. Tool Selection and Version Awareness.}
RQ2 and RQ4 are bounded by the selected tools and supported CPython versions, so
alternative decompilers or future tool versions may recover bytecode that the
evaluated tools reject or translate incorrectly. We mitigate this through
version-matched CPython environments, explicit tool-selection criteria, and
per-tool outcome reporting, which also prevents environment mismatches, loader
failures, and decompiler failures from being conflated. The results therefore
support claims about practical analyzability under the evaluated tool set and
CPython~3.8--3.14 environments rather than universal bytecode recoverability.

\vspace{3pt}
\noindent
\textit{T5. Fuzzing Inputs and Runtime Findings.}
RQ3 mutates version-matched bytecode seeds, producing malformed states that
ordinary Python source compilation may never generate, which may overrepresent
bytecode states that are not source-level reachable. The build configuration and
mutation model bound interpretation further: campaigns use honggfuzz-instrumented
CPython but not ASAN/MSAN builds~\cite{serebryany2012addresssanitizer}, and honggfuzz performs byte-level mutation of
serialized \texttt{.pyc}/marshal inputs rather than crafting valid streams. We
mitigate these threats by compiling seeds from the matching CPython unittest suite,
executing each input under the corresponding interpreter version, and treating
crashes as bytecode-level runtime findings rather than source-level
vulnerabilities, with RQ4 further bounding interpretation by testing source-recovery
reproduction. Crash reachability should therefore be read as evidence about
interpreter robustness under adversarial bytecode rather than proof of
exploitability or source-level reachability; because the builds are not
sanitizer-instrumented, the Class~B memory-corruption category is symptom-based
(fault addresses, allocator diagnostics, faulting instructions, stack context),
supporting triage prioritization but not root cause or exploitability, and
crafted-valid-bytecode payloads that import normally remain future work.

\vspace{3pt}
\noindent
\textit{T6. Fuzzing Stochasticity and Existence Scope.}
RQ3 uses one 24-hour, single-worker honggfuzz campaign per CPython version without
independent repetitions, so the 1{,}009 unique groups, per-version counts,
runtime-context ratios, and symptom-class proportions are point estimates from one
campaign set, without confidence intervals~\cite{klees2018evaluating}. We mitigate this by scoping the RQ3
claim to existence: confirmed, reproduced crash inputs demonstrate that the
bytecode execution boundary is crash-reachable, and additional runs cannot
invalidate that reachability. The counts should therefore be read as lower-bound
reachability and triage evidence rather than as stable rates or comparative
security rankings across CPython versions.

%% file: sections/concl.tex
\section{Conclusion}\label{sec:conclude}

Python bytecode is often treated as an internal cache, but this study shows that it is also a package-visible artifact, a practical analysis target, and a security-relevant interpreter input.  Across PyPI artifacts, bytecode appears in both wheels and source distributions, and some \texttt{.pyc} files lack artifact-local source within the distributed artifact.  Version-aware tools recover most modern CPython bytecode, but individual failures still require explicit accounting.  Under adversarial mutation, bytecode triggers CPython security-relevant runtime-robustness failures dominated by pointer-dereference symptoms, while selected source-recovery workflows reproduce none from ordinary Python source.  Python package security should therefore inventory bytecode directly, analyze it with version-aware tools, triage runtime findings at the bytecode level, and avoid collapsing bytecode evidence into source-level vulnerability claims.